\documentclass[aps,prb,amsmath,superscriptaddress,amssymb,longbibliography,twocolumn]{revtex4-2}
\usepackage{graphicx}
\usepackage{physics}
\usepackage{amsmath}
\usepackage{graphicx}
\usepackage{braket}
\usepackage{natbib}
\usepackage{physics}
\usepackage{color}
\usepackage{setspace} 
\usepackage[colorlinks=true,citecolor=blue,linkcolor=magenta, breaklinks=true]{hyperref}

\begin{document}
	\title{Frustrated spin-1/2 Heisenberg model on a Kagome-strip chain: Dimerization and mapping to a spin-orbital Kugel-Khomskii model}


\begin{abstract}
We investigate the quantum phases of a frustrated antiferromagnetic Heisenberg spin-1/2 model Hamiltonian on a Kagome-strip chain (KSC), a one-dimensional analogue of the Kagome lattice, and construct its phase diagram in an extended exchange parameter space. The isolated unit cell of this lattice comprises of five spin-1/2 particles, giving rise to several types of magnetic ground states in a unit cell: a spin-$3/2$ state as well as spin-$1/2$ states with and without additional degeneracies. We explore the ground state properties of the fully connected thermodynamic system using exact diagonalization and density matrix renormalization group methods, identifying several distinct quantum phases. All but one of the phases exhibit gapless spin excitations. The exception is a dimerized spin-gapped phase that covers a large part of the phase diagram and includes the uniformly exchange coupled system.
We argue that this phase can be understood by perturbing around the limit of decoupled unit cells where each unit cell has six degenerate ground states.
We use degenerate perturbation theory to obtain an effective Hamiltonian, a Kugel-Khomskii model with an anisotropic spin-one orbital degree of freedom, which helps explain the origin of dimerization.
\end{abstract}

\author{Sayan Ghosh}
\affiliation{S.N. Bose National Centre for Basic Sciences, Kolkata 700098, India.}
\author{Rajiv R. P. Singh}
\email{rrpsing@ucdavis.edu}
\affiliation{Department of Physics, University of California Davis, Davis, California 95616, USA}
\author{Manoranjan Kumar}
\email{manoranjan.kumar@bose.res.in}
\affiliation{S.N. Bose National Centre for Basic Sciences, Kolkata 700098, India.}
\date{\today}

\maketitle	

\section{Introduction}
 The possible existence of quantum spin liquids (QSL) in  frustrated magnetic materials and their applications have motivated intense research  in the  field of quantum magnetism. The QSL is a putative phase with short range magnetic order but subtle quantum orders, that has so far been elusive in real materials.  Anderson conceptualized a resonating valance bond (RVB) theory in 1973 for Mott insulating systems and the RVB state remains one the best paradigms of the QSL state \cite{ANDERSON1973153,baskaran2006resonating,balents2010spin}. In principle, the QSL state may be realized in many spin-1/2 antiferromagnetic Heisenberg  (HAF) models such as triangular \cite{chen1994optimizing,nishimori1988ground,pecher1995rvb}, Kagome \cite{mambrini2000rvb,schuch2012resonating,matan2010pinwheel,mendels2016quantum},  and ladder systems \cite{vekua2006quantum,nishiyama1996hidden,dhar2011entanglement}. 
 
 The ground states of spin-1/2 HAF model on a Kagome lattice have been found to exhibit features of a $Z_2$ spin liquid with topological order\cite{yan2011spin,mei2017gapped,depenbrock2012nature,nishimoto2013controlling}, a gapless $U(1)$ Dirac spin liquid\cite{ran2007projected,iqbal2011valence,he2017signatures,liao2017gapless}, and a valence bond crystal (VBC)\cite{skyro,rrp_vbc,hwang,ralko} with broken translational symmetry. However, the precise nature of the ground state remains inconclusive  due to the theoretical and experimental challenges in deciphering two dimensional  frustrated spin systems.   

\begin{figure}[h]
\includegraphics[width=1.0\columnwidth]{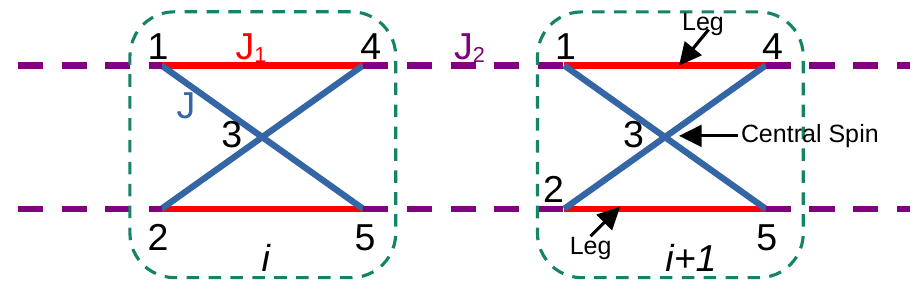}
\caption{Structure of the Kagome-strip chain. The red solid, violet broken, and blue solid lines indicate the exchange interactions $J_1$, $J_2$, and $J$, respectively. Spins in a unit cell are indicated by green dotted boxes and they are labelled $1$ through $5$. Spins labelled $1$, $2$, $4$, and $5$ are on the legs of the Kagome strip while the spins labelled $3$ form the central spins. Interactions $J_1$ and $J$ act
within a unit cell. The interactions $J_2$ link neighboring unit cells.}
\label{fig:structure}
\end{figure}

 There are several studies which focused on the simpler system of HAF spin-1/2 model on a Kagome-strip chain (KSC) to elucidate its properties. Previous studies primarily included two types of exchanges, ones along the legs of the strip ($J_1$) and ones along the rungs between a leg spin and a central spin ($J$). The structure of this strip system and exchanges corresponds to Fig. \ref{fig:structure} with $J_2=J_1$. Azaria et al \cite{P.Azaria} used various analytical approaches to show that in the limit $J/J_1\to 0$, the model has a spin excitation gap but a large number of almost zero energy singlet states. This was taken as a corroboration of one of the most striking findings of the exact diagonalization studies of the 2D Kagome antiferromagnets \cite{lecheminant,mila,waldtmann1998first}, namely, a very large number of singlet states below the lowest triplet state. Later numerical work on the Kagome-strip model using Density Matrix Renormalization Group (DMRG) \cite{pati,Steven_White_DMRG} showed that for $J_1/J=1$, {\it i.e.} a uniformly coupled Kagome strip chain, the ground states are dimerized with a spontaneously broken translational symmetry and a small spin-gap.

 In 2016 a new class of materials, A$_2$Cu$_5$(TeO$_3$)(SO$_4$)$_3$(OH)$_4$ (where $A=Na,K$)\cite{tang2016synthesis} were synthesized whose effective  spin placements resembles the  kagome strip chain structure shown in Fig. \ref{fig:structure}. Interestingly, Cu atoms distances in A$_2$Cu$_5$(TeO$_3$)(SO$_4$)$_3$(OH)$_4$ (where A=Na,K)  are not the same  \cite{tang2016synthesis}, therefore, we expect more non-uniformity in the exchange interactions leading to the non-uniform KSC ladder. This work motivated further theoretical studies of a model with three type of exchanges $J$, $J_1$ and $J_2$ on a non-uniform KSC ladder system to understand the exotic phases arising in this system. In particular, in presence of an external magnetic field, up to fifteen distinct magnetization plateau phases were found on kagome strip chain\cite{morita2021magnetic,morita2018magnetization} for various parameters, and these plateaus of kagome strip chain were not always the same as the plateaus of the 2D Kagome lattice\cite{hida2001magnetization,sakai2011critical,capponi2013numerical,nishimoto2013controlling,nakano2015magnetization,picot2016spin,schnack2020magnon}. 
 The 3/5 plateau ground state of the KSC  was argued to map on to a spin-$1$ Haldane phase\cite{morita2021magnetic}, while a gapless spin liquid state with doubly degenerate entanglement spectra due to resonating dimer-monomer structures was observed at the $1/5$ magnetization plateau\cite{morita2021resonating}. However, the understanding of the ground state properties of the spin-1/2 antiferromagnet on kagome strip chain is still incomplete, and the mechanism of ground state dimerization, phase diagram, and the nature of the singlet and triplet excitation spectra are not fully understood.
 
In this paper, we begin by analysing the ground state properties of a five-spin unit cell. They consist of a S=3/2 state for $0<J_1/J\leq 0.5$,  two degenerate sets of $S=1/2 $ ground states or spin-half states with orbital degree of freedom (SHSODF) for $0.5<J_1/J<1$,  at $J_1/J=1$ there are six degenerate ground states which can be considered as a spin-half system with 3-state orbital degree of freedom, and for $J_1/J>1$ the ground states consist of a free spin-1/2 with no additional degeneracies. In presence of inter-unit exchange coupling $J_2$ we  construct a comprehensive quantum phase diagram (QPD) in the exchange-parameter space. Our analysis identifies five distinct ground state phases, including two symmetry-broken phases characterized by two fold degenerate ground states. We also derive effective Hamiltonian models using degenerate perturbation theory. For $J_1=J$ the system maps on to a one-dimensional Kugel-Khomskii (KK) model with a highly anisotropic spin-one orbital degree of freedom that provides direct insights into the mechanism for the origin of dimerization. 

The paper is divided into five sections. The model Hamiltonian and the numerical methods are discussed in section II. Section III covers the quantum states  of the unit cell and  also the results of the computational studies leading to the identification of different phases and phase boundaries. In the fourth section  we use perturbation theory to obtain and study  the effective Hamiltonian of coupled spin and orbital degrees of freedom. The final section summarizes our main findings.

\section{Model and Methods}
We consider an isotropic antiferromagnetic Heisenberg model on the kagome strip chain shown in Fig.\ref{fig:structure}:
\begin{multline} \label{eq1}
H=\sum_{i}J_1(\Vec{S}_{i,1}\cdot\Vec{S}_{i,4}+\Vec{S}_{i,2}\cdot\Vec{S}_{i,5})+J(\Vec{S}_{i,1}\cdot\Vec{S}_{i,3}+\Vec{S}_{i,2}\cdot\Vec{S}_{i,3}\\+\Vec{S}_{i,3}\cdot\Vec{S}_{i,4}+\Vec{S}_{i,3}\cdot\Vec{S}_{i,5}) + J_2(\Vec{S}_{i,4}\cdot\Vec{S}_{i+1,1}+\Vec{S}_{i,5}\cdot\Vec{S}_{i+1,2})
\end{multline}
Here, $S_{i,x}$ represents spin at site $x$ of the $i^{th}$ unit cell. $J$ represents the spin exchange with the central spin, whereas $J_1$ and $J_2$ are two alternate exchange couplings along the legs. $J_1$ acts between the spins within the unit cell while neighbouring spins from two nearest unit cells interact through exchange $J_2$ as shown in the Fig.\ref{fig:structure}. We will set the energy scale by working with $J=1$. We will take all interactions to be antiferromagnetic ($J_i>0$).

The model Hamiltonian is many body in nature and the Hilbert space dimension increases as $2^N$ for N spins. Therefore, we can employ the Exact Diagonalization (ED) method only for small system sizes, and Density Matrix Renormalization Group(DMRG)\cite{Steven_White_DMRG,Schollwock_DMRG,KA_Hallberg_DMRG} is used for larger system sizes up to $32$ unit cells (160 spins) with periodic boundary condition (PBC). ED can provide qualitative answers related to local properties, however, one requires larger system sizes to give reliable phase boundaries in the thermodynamic limit. The DMRG method is a state of the art numerical method to find the ground state and low-lying excited state properties of quasi one-dimensional systems. In the DMRG calculations, we retain up to $800$ density matrix eigenstates and the truncation error is kept at less than $10^{-8}$ to ensure reliable accuracy of the results. Maximum  error in the energy is less than 0.1 \%.  

We calculate some low lying energy gaps such as first singlet excited state gap $E_\sigma$, second singlet excited state gap $E_{2\sigma}$ and lowest triplet excited state gap $E_m $ which are defined as   : 
\begin{subequations}\label{eq2}
\begin{align} 
E_\sigma &= E_1 (S=0) - E_0 (S=0), \label{eq2a}\\
E_{2\sigma} &= E_2 (S=0) - E_0 (S=0), \label{eq2b}\\ 
E_m &= E_0 (S=1) - E_0 (S=0), \label{eq2c}
\end{align}
\end{subequations}
where $E_0 (S=0) , E_1 (S=0), E_2 (S=0)$ and $E_0 (S=1)$ are energies of ground state, first excited singlet state, second excited singlet state and lowest triplet state respectively. Crossings of these excited state energy levels and their symmetries can be used to determine the ground state phase boundaries.
Another useful quantity is the ground state energy curvature or its second derivative $E^{\prime \prime}$ in parameter space defined as
\begin{equation}
E^{\prime \prime}=-\frac{1}{N}\frac{d^2E}{dJ_\alpha^2},
\end{equation}
where $E$ is the ground state energy and $J_{\alpha}$ is an exchange parameter. This second derivative of the ground state energy is analogous to the heat capacity at finite temperatures. The peaks in $E^{\prime \prime}$ indicate a phase transition. 

The spin-spin correlation functions in the ground state are defined as, 
\begin{equation}
C(r=|j-i|) = \bra{\psi_{0}} S_i\cdot S_{j} \ket{\psi_{0}}, 
\end{equation}
where $\ket{\psi_0}$ is the ground state and $r=|j-i|$ represents the distance between the reference site $i$ and site $j$ as shown in supplementary materials. The correlation function provides important information on the spin arrangement and range of magnetic order in the state. The nearest neighbour correlation $C(|i-j|=1)$ is also denoted as $B_a$ for bond $a$. The product of $B_a$ and the corresponding exchange on bond $a$ is the local bond energy of the given state.

In some part of the parameter space of the model, the ground state is doubly degenerate in the thermodynamic limit. In the doubly degenerate ground states $\ket{\phi_{\alpha}}$ and $\ket{\phi_{\beta}}$ we can define a $2\cross2$ matrix with elements $B_{ij}^{\alpha\beta}$.  
\begin{equation}
B_{ij}^{\alpha\beta} = \bra{\phi_{\alpha}}   S_i\cdot S_{j}  \ket{\phi_{\beta}} 
\end{equation}
For a large enough system $B_{ij}^{\alpha\alpha} = B_{ij}^{\beta\beta}$ and the two broken symmetry states have correlation functions $B_{ij} = B_{ij}^{\alpha\alpha} \pm B_{ij}^{\alpha\beta}$ .
In a finite system with periodic boundary condition these two states are not exactly degenerate but form the two lowest eigenstates separated from the rest. In this case $B_{ij}^{\alpha\alpha}$ is not exactly equal to $B_{ij}^{\beta\beta}$. The matrix elements in these two low lying states depend weakly on the system sizes.  
We define $B_{ij}$ in a finite system as \cite{schollwock2008quantum,mk_soos}:
\begin{equation}\label{eq3}
  B_{ij} = \frac{B_{ij}^{\alpha\alpha} + B_{ij}^{\beta\beta}}{2} \pm B_{ij}^{\alpha\beta}. \\
\end{equation} 
The pattern of $B_{ij}$ on various bonds of the lattice provides a direct visualization of the nature of order. The matrix element $B_{ij}^{\alpha\beta}$ is called dimer order. The bond order pattern in the two broken symmetry states are related by spatial symmetry. The finite value of the dimer order parameter is a characteristic of the dimerized ground states.


\section{Results and Quantum phase diagram}
In this section we discuss the ground state properties in various parameter regimes. The kagome strip chain ladder is composed of unit cells of five spin-$1/2$ spins as shown in Fig.\ref{fig:structure} and  competing exchanges $J$ and $J_1$ within the unit cells give rise to frustration which leads to various types of spin states.  We first analyze the ground states of a single unit cell in various limits of $J_1/J$.  Thereupon, we study the effect of inter unit-cell exchange $J_2$ and quantum phases in the ground state in various parameter regimes.  We find that five  quantum phases  arise by tuning the parameters $J_1$  and $J_2$ . The quantum phases are discussed first and, thereafter,  we discuss  the quantum phase diagram and  the phase boundaries based on the  energy level crossovers, bond orders and spin correlation functions. 

As seen in Fig.~\ref{fig:structure} a unit cell has two corner sharing triangles. Varying $J_1/J$ leads to a net spin-$1/2$ or spin-$3/2$ ground state in a unit cell with or without additional degeneracies, as shown below. When we find additional ground state degeneracies other than those associated with the $SU(2)$ symmetry of the spin, we call it an orbital degree of freedom inside a unit cell in analogy with the Kugel-Khomskii model. Later, we show using degenerate perturbation theory that the model in Eq.\ref{eq1} can be mapped on to an interacting Hamiltonian of spins and orbitals. For $J_1/J=1$ the effective model Hamiltonian in Eq. \ref{eq1} is a 1D Kugel-Khomskii model with a 3-state or anisotropic spin-one orbital degree of freedom \cite{kugel1973crystal,li19984,itoi1999phase,frischmuth1999thermodynamics,pati1998alternating,azaria1999one,zasinas2001phase}. 

\subsection{Quantum states of a unit cell}\label{qsu}
We consider a single unit cell and omit the index of the unit cell, the isotropic Heisenberg Hamiltonian of the unit cell can be written as 
\begin{equation} \label{eq4}
H_{uc}=J\Vec{S}_3\cdot(\Vec{S}_1+\Vec{S}_2+\Vec{S}_4+\Vec{S}_5)+J_1(\Vec{S}_1\cdot\Vec{S}_4+\Vec{S}_2\cdot\Vec{S}_5). 
\end{equation}
Here, $\Vec{S}_i$, represent the spin vector for site $i$ for the five sites shown in Fig. \ref{fig:structure}).  We can find all the states by adding spins in the order: $\Vec{S}_{14}=\Vec{S}_1+\Vec{S}_4$, $\Vec{S}_{25}=\Vec{S}_2+\Vec{S}_5$, $\Vec{S}_{1245}=\Vec{S}_{14}+\Vec{S}_{25}$, and the total spin $\Vec{S}_t=\Vec{S}_{12345}=\Vec{S}_{1245}+\Vec{S}_3$. The Hamiltonian in Eq.\ref{eq4} can be rewritten as,
\begin{equation} \label{eq5}
\begin{split}
H=\frac{J}{2}(\Vec{S}_{t}^2-\Vec{S}_{1245}^2-\vec{S_3}^2)+\frac{J_1}{2}(\vec{S}_{14}^2+\vec{S}_{25}^2 \\
-\vec{S}_1^2-\vec{S}_2^2-\vec{S}_4^2-\vec{S}_5^2)    
\end{split}
\end{equation}
Since the square of the spins $\Vec{S}_{14}$, $\Vec{S}_{25}$, $\Vec{S}_{1245}$, and $\Vec{S}_{t}$ are conserved quantities, the energy eigenvalues can be obtained by replacing each $S^2$ operator by its eigenvalue $s(s+1)$.

\begin{figure}[h]
\includegraphics[width=1.0\columnwidth]{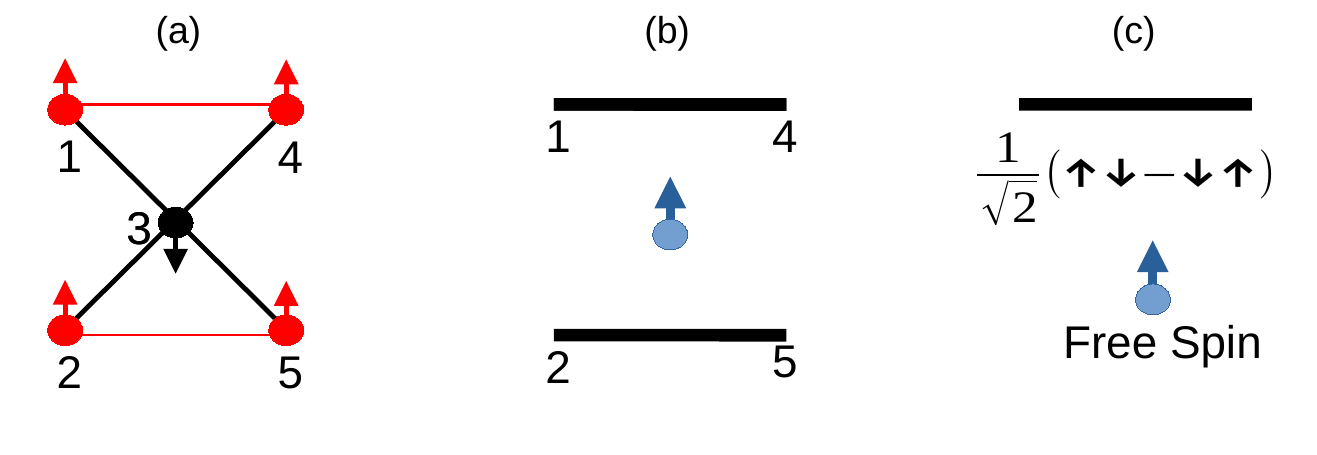}
\includegraphics[width=1.0\columnwidth]{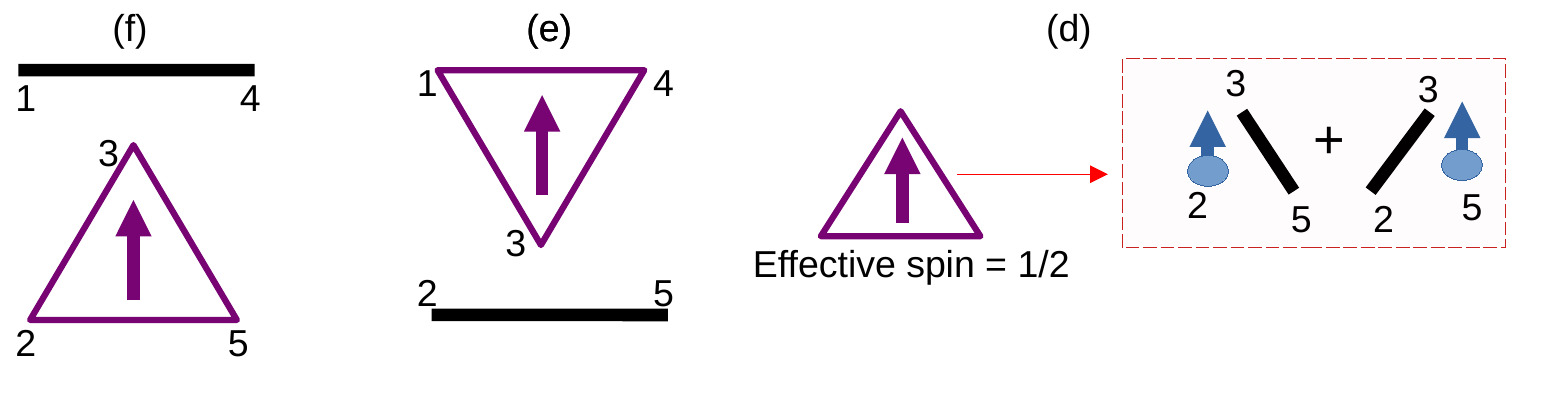}
\caption{Ground states configurations of a unit cell. (a) For $0< J_1/J < 1/2$, the effective spin is $3/2$, all leg spins are ferromagnetically aligned while the central spin couples antiferromagnetically to the leg spins. (b) For $J_1>J$, effective spin is $1/2$ which resides at central site and upper and lower leg spins form singlet dimers. (c) Singlet dimer and free spin are shown as solid black line and up arrow with solid blue circle respectively. (d) The triangle with arrow represents a spin-half state formed from 3 spins that is also an equal superposition of two states, each consisting of a singlet dimer and a free spin-$1/2$. (e) and (f) Two degenerate set of spin-half ground states arise for $1/2<J_1/J<1$ with a dimer on one leg and a triangle with arrow. }
\label{fig:unit_cell_phases}
\end{figure}

Different combinations of allowed $s_{t}$, $s_{1245}$, $s_{14}$, and $s_{25}$ yield all 32 states of the unit cell as shown in the table in Supplementary Materials.
In the various $J_1/J$ limits, the ground states of the unit cell are of three types:

(i) Spin $3/2$ state: For $J_1/J\leq 0.5$, the energy is minimized by setting $S_{14}=S_{25}=1$, $S_{1245}=2$, and $S_{t}=3/2$ and the  ground state energy $E_1=\frac{-3J}{2}+\frac{J_1}{2}$. These states have all four leg spins aligned ferromagnetically and these combine antiferromagnetically with the central spin to minimize the dominant energy $J$ (See Fig.\ref{fig:unit_cell_phases}(a)). 

(ii) Spin-$1/2$ states with orbital degree of freedom (SHSODF): For  $1/2<J_1<1$, an intriguing feature emerges in the unit cell. The ground states in this parameter regime have effective spin $S=1/2$ but it is four fold degenerate.

Only two fold degeneracy comes from spin-$1/2$ and other two is due to resonating structure of the system shown in Fig.\ref{fig:unit_cell_phases}(c,d,e,f). The ground state energy is  $E_2=-J-\frac{J_1}{2}$, obtained by setting $s_{14}=0$, $s_{25}=1$, $s_{1245}=1$, $s_{t}=1/2$, or $s_{14}=1$, $s_{25}=0$, $s_{1245}=1$, $s_{t}=1/2$. In either case, one of the leg spin-pair, either upper or lower, forms a singlet while the other leg spin-pair forms a triplet which combines with the middle spin-$1/2$ to form a resonating structure with one singlet and one free spin as shown in Fig.\ref{fig:unit_cell_phases}(d).  
Interestingly, for $J_1 = 1$, the ground state of unit cell is six fold degenerate signifying $3$ orbital states.

\begin{figure}[h]
\includegraphics[width=1.0\columnwidth]{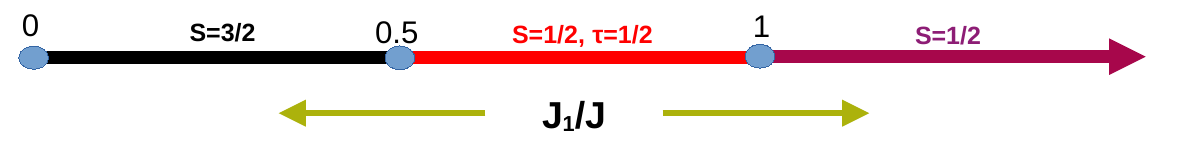}
\caption{Phase Diagram for a single unit cell. Three distinct phases emerge at different regimes of $J_1/J$ - (i) Spin-3/2 state at $0<J_1/J<0.5$, (ii) Spin-$1/2$ with orbital degree of freedom (SHSODF) at $0.5<J_1/J<1$ and (iii) Spin-1/2 with dimerized leg at $J_1/J>1$.}
\label{fig:phase_diagram_EE}
\end{figure}

(iii) $S=1/2$ with dimerized legs: For $J_1>J$, both leg states of the pentagon form strong singlet bonds and there is a free spin-1/2 at the centre as shown in the Fig.\ref{fig:unit_cell_phases}(b). The ground state energy $E_3=\frac{-3J_1}{2}$ is obtained with quantum numbers $s_{14}=s_{25}=0$, $s_{1245}=0$, and $s_{t}=1/2$. 

Now let us analyze the effect the inter unit cell spin exchange $J_2$ has on the various states and the evolution of various type of quantum phase boundaries. 

\subsection{Quantum phases on the ladder}\label{QPD}

In the Hamiltonian in Eq. \ref{eq1}, $J_1/J$ determines the ground state within a unit cell. For small $J_2/J$, one can use degenerate perturbation theory to study properties of the system. These will be discussed later. However, when $J_2$ and $J_1$ are comparable, the perturbative approach breaks down. The full evolution of the phases and phase boundaries requires a computational approach. Phase boundaries can be determined using quantum numbers of low energy excitations, second derivative of the ground state energy, bond order calculations, and spin-spin correlation functions.

An important quantity is the quantum number of the lowest energy excitation. The Lieb-Schultz-Mattis (LSM) theorem says that the for half-integer spins in a unit cell, in the thermodynamic limit, either the lowest spinful excitation energy is gapless or there must be degenerate ground states \cite{lieb1961two,tasaki2022lieb}. However, in a finite system, the excitation gap above a non-degenerate ground state can be finite and the lowest excited state could be singlet or triplet. On tuning the competing  exchanges the higher lying states may come down to become the lowest excited state. The crossing of low lying states can itself provide accurate measures of phase boundaries. When the ground state is degenerate, it may indicate a broken symmetry state and the dimer order $B^{01}_{ij}$ parameter, discussed earlier, is constructed. 

In this section, we first introduce the various quantum phases, and later, the phase boundaries will be discussed. There are five phases in the $J_1$ and $J_2$ parameter space, but two of these phase-I and II, are related by a crossover. The basic properties of these phases are given below: \\

\begin{figure}[h]
\includegraphics[width=1.0\columnwidth]{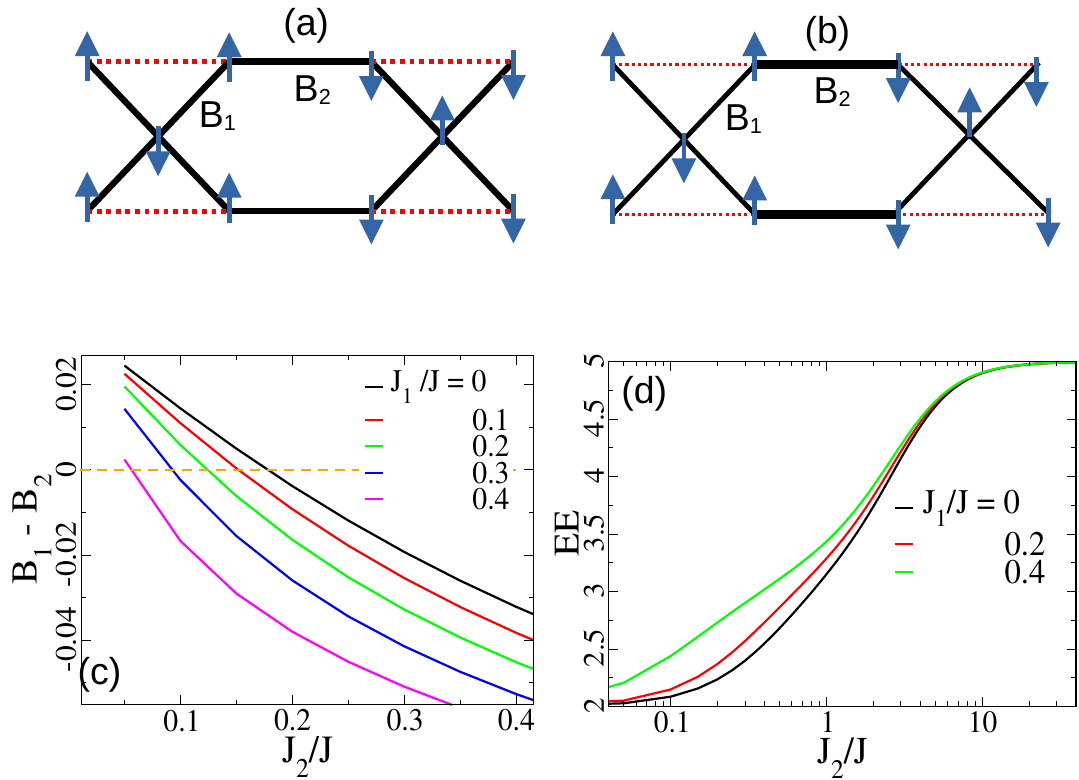}
\caption{The line widths in (a) and (b) depict the strength of spin correlations on different bonds. Black (red) solid (dotted) lines correspond to anti-ferromagnetic (ferromagnetic) correlations respectively. (a) $J_1/J = 0.1$ and $J_2 = 0.15$ in phase-I, and (b) $J_1/J = 0.1$ and $J_2 = 1.8$ in phase-II. (c) The difference $B_1 - B_2$ is shown as a function of $J_2/J$ for several $J_1/J$. For every $J_1/J$ up to $1/2$, $B_1 - B_2$ starts from a positive value, crosses zero, and ends at a negative value. The $J_2/J$ at which $B_1 - B_2$ crosses zero is taken as the crossover point from phase-I to phase-II in the phase diagram (see Fig.\ref{fig:phase_diagram}). (d) Variation of entanglement entropy (EE) in units of $\ln(2)$ with $J_2/J$ for various values of $J_1/J$ in phase-I and II. At large $J_2/J$, the EE approaches $5\ln{2}$, indicating that each unit cell can access $2^5$ states.}
\label{fig:phase_I_II}
\end{figure}

(i,ii) \textbf{Phase-I and II :}  The ground states of both phases are singlets and the schematic representation of local spin arrangements in phase-I and II are shown in Fig.~\ref{fig:phase_I_II}(a),(b). phase-I corresponds to small $J_2$, where perturbatively the system maps on to a spin-$3/2$ chain. The spins on the legs $2-5$ and $1-4$ are aligned parallel to each other, whereas the central spin is aligned antiferromagnetically to the leg spins. As $J_2$ increases the inter-dimer correlations gradually become stronger. In phase-I, the nearest-neighbor spin correlations between central and leg spins, $B_1=<S_{3,i}\cdot S_{4,i}>$, is stronger or comparable to that between two leg spins of adjacent unit cells, $B_2=<S_{4,i}\cdot S_{1,i+1}>$, whereas in phase-II the correlation $B_2$ is stronger compared to $B_1$. We define the crossover point between phase-I and phase-II as one where $B_1$ equals $B_2$. The difference $B_1-B_2$ is shown as function of $J_2$ in Fig.\ref{fig:phase_I_II}(c). 

As the leg spins are gradually tied into singlets with spins in neighboring unit cells with increasing $J_2$, it leaves only central spin-$1/2$ within each unit cell. This can be regarded as a gradual crossover from a spin-$3/2$ chain to a spin-$1/2$ chain. However, both nearest-neighbor spin-$3/2$ chain and spin-$1/2$ chain have been argued to have the same long distance behavior \cite{K.Hallberg}. Thus one does not expect a sharp transition between the two phases, only a crossover. The correlation between the central spins, $C(r) = <\vec{S}_i \cdot \vec{S}_{i+r}>$ follows an algebraic decay as a function of distance in both phases. The correlations between the leg spins decays algebraically in phase-I whereas it shows short range correlations in phase-II (See Supplementary Materials). The spin-correlations are consistent with the system belonging to the half integer spin chain universality class, characterized  by a conformal central charge, $c=1$, with a $1/r$ decay of correlations with logarithmic corrections \cite{K.Hallberg}. 

In both phase-I and II the lowest excitation is triplet, which becomes gapless in the thermodynamic limit as expected for half-integer spin-chains. We also calculate the bipartite entanglement entropy (EE) of a single unit cell with respect to the rest of the system. In the small $J_2$ limit the effective spin of the unit cell is $3/2$ and only four states participate in the ground state, whereas in the large $J_2$ limit the intra-unit-cell couplings become irrelevant and each unit cell is able to explore all $2^5$ states, and each spin in a unit cell entangles independently with spins in other unit cells. Thus the bipartite entanglement entropy of a single unit cell in the ground state increases gradually from $2\ln{2}$ to $5\ln{2}$ as $J_2$ increases from $0$ to $\infty$ (Fig. \ref{fig:phase_I_II}(d)). 

\begin{figure}[h]
\includegraphics[width=1.0\columnwidth]{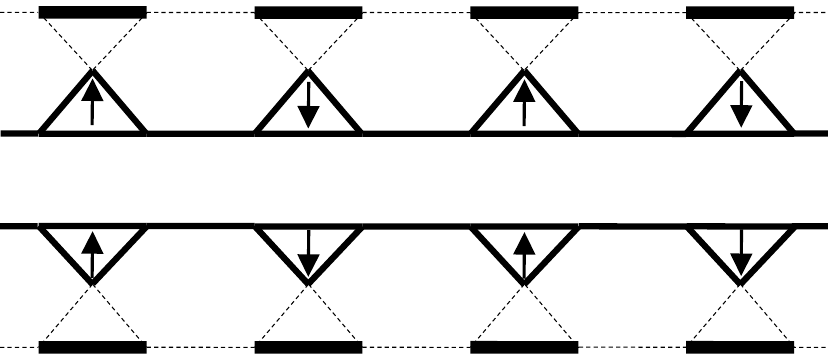}
\caption{Schemetaic representation of the two degenerate ground states in phase-III. Solid line and triangle with arrow represent singlet dimer and the effective spin-$1/2$ as in Fig.\ref{fig:unit_cell_phases}(c,d). Spin pairs on upper (or lower) leg are tied in intra unit-cell singlets while the remaining spins form a spin-half chain. These two states are related by reflection symmetry.}
\label{fig:local bond strength 1}
\end{figure}

(iii) \textbf{Phase-III:} This phase arises for $1/2 < J_1/J < 1$ and small $J_2$. 
For $J_2 = 0$ and $1/2 < J_1/J < 1$, decoupled unit cell ground state is of type (ii) described in previous section (shown in Fig.~\ref{fig:unit_cell_phases}(d,e)). The spin-1/2 kagome strip chain model in Eq.\ref{eq1}, in the weak $J_2$ limit, can be mapped to an anisotropic Kugel-Khomskii Model which will be discussed in the next section in more detail.
In this phase, either the upper (or lower) leg spin pairs form intra-unit cell singlets, whereas, the other three spins in a unit cell form an effective spin-1/2 which are coupled from cell to cell into a spin-half chain. The ground state is doubly degenerate and these two states are related by reflection symmetry about the axis passing through central spins. Both  degenerate ground states consist of an effective isotropic Heisenberg chain as shown in Fig.\ref{fig:local bond strength 1}.  
       
\begin{figure}[h]
\includegraphics[width=1.0\columnwidth]{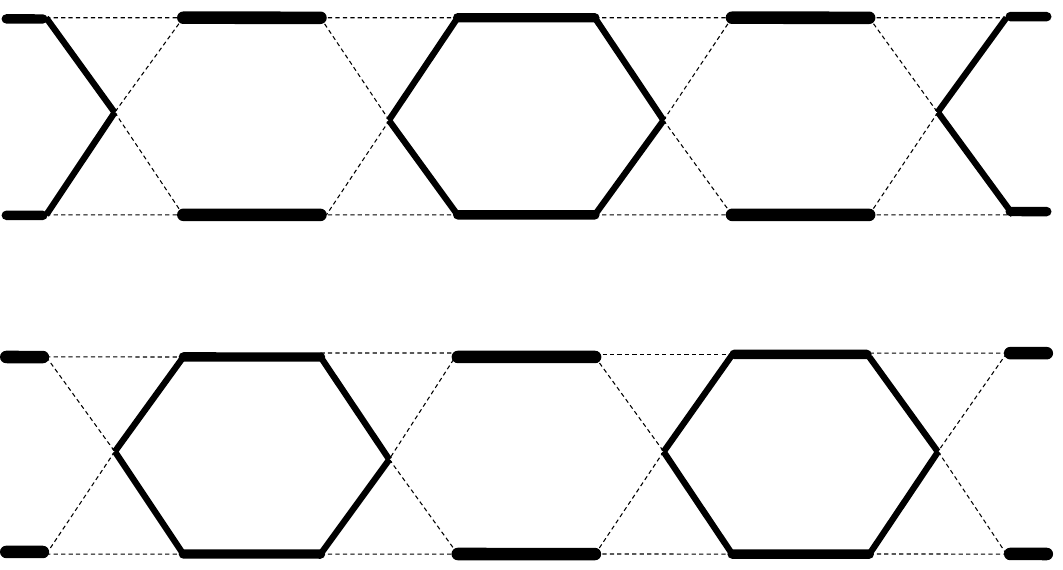}
\caption{Schematic representation of two degenerate ground state in phase-IV, consisting of alternating dimer and hexamer pattern for $J_2/J=J_1/J$. These two states are related by translation.}
\label{fig:local bond strength 2}
\end{figure}

(iv) \textbf{Phase-IV:} This phase exhibits the formation of weakly interacting alternating dimer and hexamer (hexagonal structure) pair structures shown in Fig.\ref{fig:local bond strength 2}, and the ground state is a doubly degenerate singlet in the thermodynamic limit, similar to the doubly degenerate dimerized ground state of the Majumdar-Ghosh model\cite{majumdar1969next}. The expectation value of the bond order $B_{ij}$ in Eq.\ref{eq3} is shown in Fig.\ref{fig:local bond strength 2} and is clearly visible in phase-IV, which agrees with the DMRG calculation for uniform exchange model done previously for $J=J_1=J_2=1$\cite{RRP_Singh}. The spin correlation in this phase decays exponentially and the singlet-triplet gap is finite. We find that in some parameter ranges, in this phase, there are a large number of low-lying singlet states below the lowest triplet excitation for finite system.      

\begin{figure}[h]
\includegraphics[width=1.0\columnwidth]{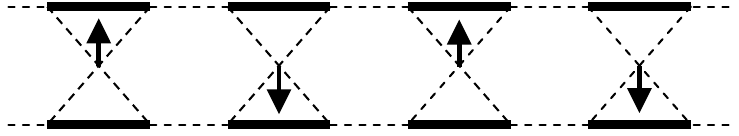}
\caption{Schematic representation of phase-V, where the leg spins form intra unit-cell singlets while the central spins form an antiferromagnetic chain.}
\label{fig:phase_V}
\end{figure}

(v) \textbf{Phase-V :} The ground state of this phase shows antiferromagnetic arrangement of spins along both leg and central spins, however, there is quasi-long range order for the central spins and only short-ranged correlations for the leg spins. The short range order along the legs is due to strong singlet formation inside the 
unit cell as shown in Fig.\ref{fig:phase_V}. This phase arises for $J_1/J >1$. The central spins of the unit cells (See Fig.\ref{fig:unit_cell_phases}(b)) give rise to a weakly coupled spin-half Heisenberg chain. 

\subsection{Quantum phase boundary}
\begin{figure}[h]
\includegraphics[width=1.0\columnwidth]{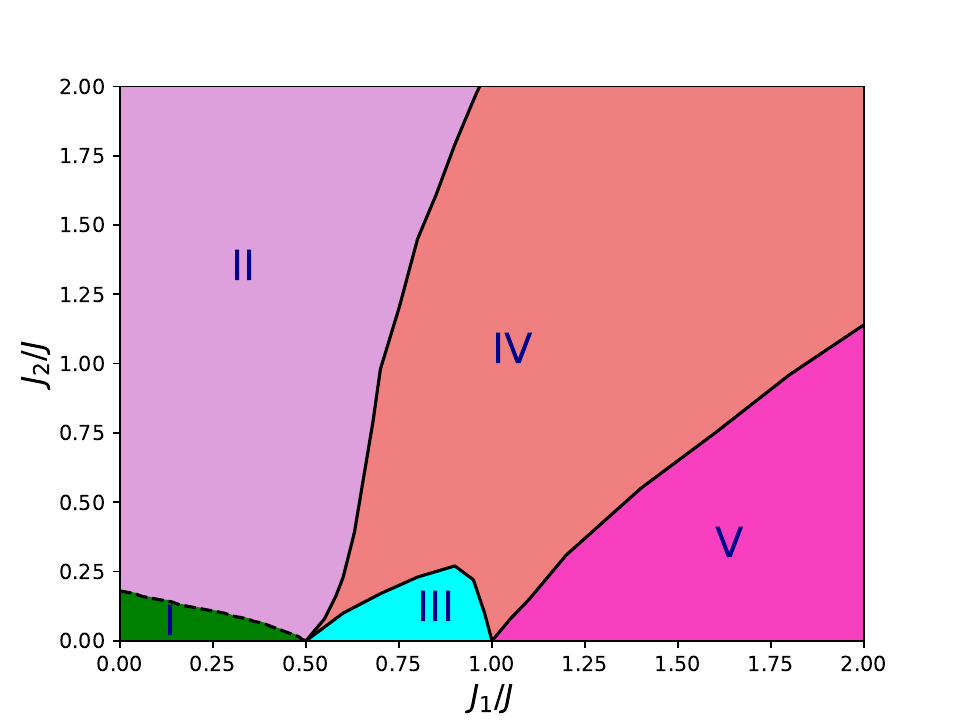}
\caption{Quantum phase diagram and its phase boundaries in the $J_1/J$-$J_2/J$ plane. The dashed line represents a crossover between phase-I and II. Solid black lines represent phase boundaries between phase-III and IV, phase-II and IV, and phase-IV and V.
} 
\label{fig:phase_diagram}
\end{figure}
The quantum phase diagram is shown in Fig. \ref{fig:phase_diagram}. 
All phases in the phase diagram are consistent with the LSM theorem \cite{lieb1961two,tasaki2022lieb}, namely, as translational symmetry along the chains is present in all phases except phase-IV, they all have gapless spin-excitations (in phase-I, II, III and V). On the other hand, phase-IV has a doubly degenerate ground state with broken translational symmetry. 

We primarily rely on the crossing of the two lowest singlet excitation-gaps $E_{\sigma}$ and the lowest triplet excitation-gap  $E_{m}$ , defined in Eq. \ref{eq2}, to determine the phase boundaries. Similar to LSM theorem,  Okamoto-Nomura pointed out that in 1D systems with spin-rotational symmetry, if the lowest singlet excitation is lower than the lowest triplet excitation in a finite system, the ground state becomes doubly degenerate, whereas if the triplet excitation is lower it indicates a gapless ground state with quasi-long range spin order \cite{okamoto1992fluid} and  this criterion remains valid in other low -dimensional systems \cite{giamarchi2003quantum,nomura1993phase,sandvik1998critical,dagotto1996surprises}. Therefore, lowest excited state crossings can be a good criteria for determining the phase boundary and it can be further supported by calculations of energy curvature and spin correlations.

The boundary of phase-I and II is a crossover and it is determined by using the correlation difference $B_1-B_2 $ as shown in the Fig.\ref{fig:phase_I_II}. In phase-I the intra unit cell correlation $B_1$ is stronger compared to inter unit-cell correlation $B_2$, whereas $B_2$ is stronger in phase-II. The crossover happens at $B_1=B_2$.   

\begin{figure}[h]
\includegraphics[width=1.0\columnwidth]{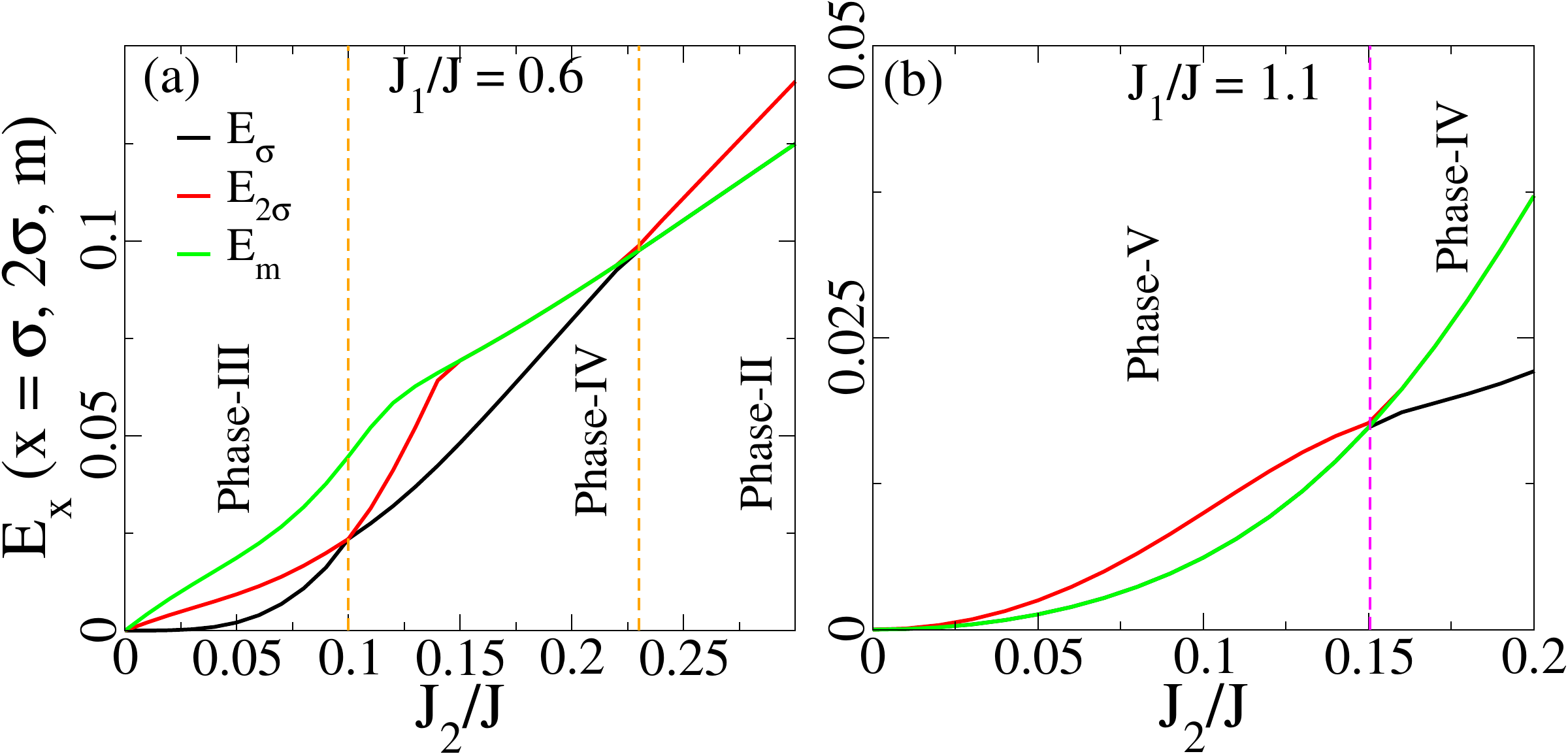}
\caption{ Excited state energy gaps $E_x$ (where $x=\sigma,m$ represent singlet and triplet states respectively) versus $\frac{J_2}{J}$ for (a) $\frac{J_1}{J} = 0.6$ and
(b) $\frac{J_1}{J} = 1.1$. (a) Two different singlet excitations, reflecting different broken symmetries, cross at the first phase boundary from phase-III to phase-IV at $J_2/J=0.1$. Above $J_2/J=0.23$, a triplet becomes the lowest excitation denoting a transition from phase-IV to phase-II. (b) Below $J_2/J=0.15$ a triplet is the lowest excitation in phase-V. Beyond $J_2/J=0.15$, a singlet becomes the lowest excitation denoting a transition to phase-IV.} 
\label{fig:energy_level_crossover}
\end{figure}

We determine phase-III to phase-IV transition by the crossing of two different singlet states and the transition from either phase-II or phase-V to phase-IV by the crossing of singlet and triplet excitations. As both phase-III and phase-IV have two degenerate singlet ground states in the thermodynamic limit but they reflect different broken symmetries, the corresponding first excited states must be singlets but of different symmetry. In Fig.\ref{fig:energy_level_crossover}(a,b), $E_\sigma$, $E_m$  and  $E_{2\sigma}$ are plotted as a function of $J_2/J$ for two different values of $J_1/J$. For $J_1/J=0.6$, the crossing between two singlet excitations occurs at $J_2/J=0.1$  and the phase boundary between phase-III and phase-IV  is drawn using this criteria (See Fig.~\ref{fig:energy_level_crossover}(a)). The crossing between $E_\sigma$ and $E_m$ occurs at $J_2/J_1=0.23$ (See Fig.~\ref{fig:energy_level_crossover}(a)) and that determines the phase boundary between phase-IV and phase-II. 
For  $J_1/J = 1.1$,  the triplet and singlet excitations cross at $J_2/J =0.15 $ denoting a transition from phase-V to phase-IV.

\begin{figure}[h]
\includegraphics[width=1.0\columnwidth]{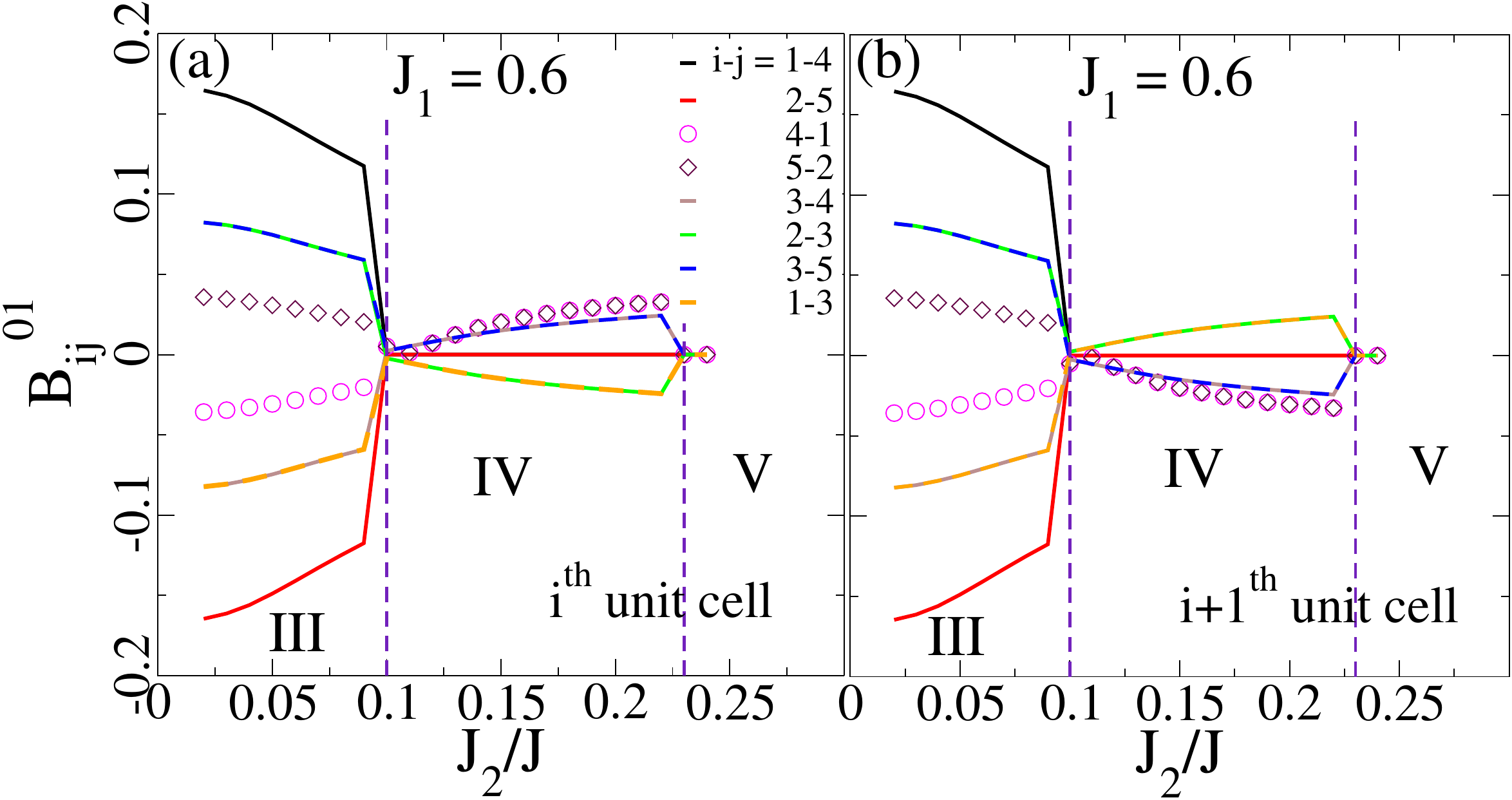}
\caption{Dimer order $B_{ij}^{01}$ (See Eq.~\ref{eq3}) as a function of $J_2/J$ for various neighboring spin-pairs of (a) $i^{th}$ unit cell and (b) $(i+1)^{th}$ unit cell for $J_1/J = 0.6$. The square and circle symbols represent the dimer order of the inter-unit cell bond, corresponding to the two bonds between the $i^{th}$ and $(i+1)^{th}$ unit cells. $B^{01}$ are finite for all spin-pairs for $J_2/J <0.1$ but for $J_2/J>0.1$, $B_{14}^{01}$ and $B_{25}^{01}$ vanish, while others remain finite. 
For $J_2/J>0.23$ a triplet becomes the lowest excitation (shown in Fig.\ref{fig:energy_level_crossover}) and $B_{ij}^{01}$ vanish for all spin-pairs.}
\label{fig:bond_order}
\end{figure}

These phase boundaries can be further supported by calculations of dimer order $B_{ij}^{01}$ (See Eq.~\ref{eq3}). Fig.~\ref{fig:bond_order} shows the changes in dimer order as a function of $J_2/J$ for a particular value of $J_1/J = 0.6$. It changes dramatically at the different phase boundaries. When the triplet is the lowest excitation, the dimer order vanishes. When non-zero the dimer order highlights the ordering patterns which arise from distinct broken symmetries (as shown for phase-III and IV in Fig. \ref{fig:local bond strength 1} and \ref{fig:local bond strength 2}, respectively).

\begin{figure}[h]
\includegraphics[width=1.0\columnwidth]{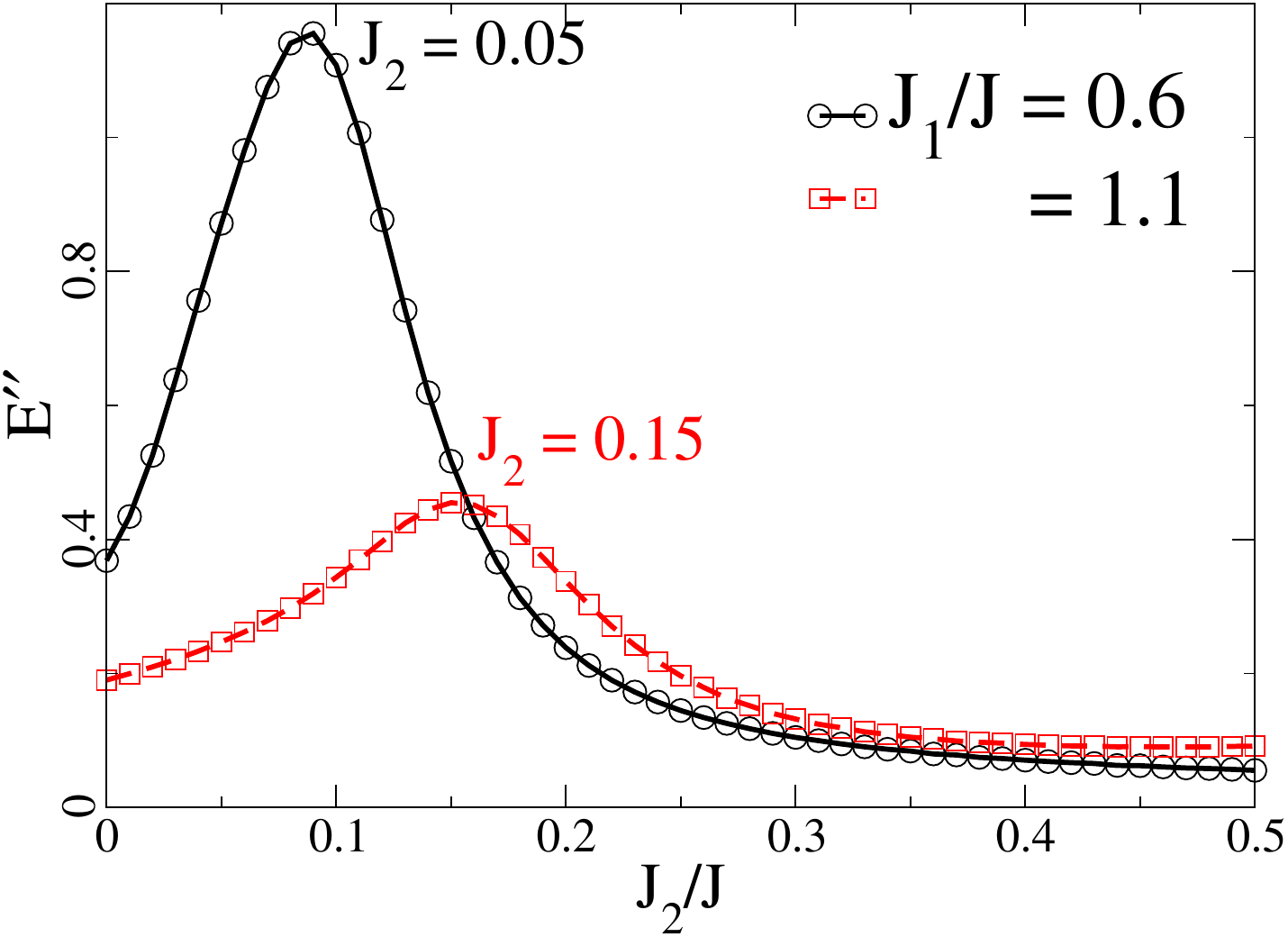}
\caption{Second derivative of ground state energy ($E^{\prime\prime}$) with respect to $J_2/J$ for a given $J_1/J$. For $J_1/J = 0.6$ (black solid) and $1.1$ (red dashed) the peaks are at $J_2/J=0.09$ and $0.15$ respectively.} 
\label{fig:energy_derivative}
\end{figure}

One can also obtain the phase boundaries using the energy curvature $E^{\prime\prime}$. In Fig.\ref{fig:energy_derivative}, the peak in $E^{\prime\prime}$ are shown by arrows. The peaks in red dashed and black solid curves in Fig.\ref{fig:energy_derivative} are close to the phase boundaries between phase-IV and phase-V and between phase-III and phase-IV, respectively. However, they are less reliable than the excited state crossings in determining the phase boundary.


\section{Degenerate Perturbation Theory: Mapping onto Kugel-Khomskii Models}
In this section we consider degenerate perturbation theory around the limit of decoupled unit cells. When ground states have no additional degeneracies except those related to $SU(2)$ symmetry, the low energy Hamiltonian is a nearest-neighbor Heisenberg model.
For $J_1/J<1/2$, one obtains a spin-$3/2$ Heisenberg model and for $J_1/J>1$ one obtains a spin-half Heisenberg model. We will focus attention on $1/2 < J_1/J \leq 1$  where the unit cell ground states have spin-$1/2$ but have additional degeneracies. We call these additional degeneracies orbital degree of freedom in analogy with Kugel-Khomskii models and use degenerate perturbation theory to obtain effective low energy Hamiltonians.

Inside phase-III, {\it i.e.} for $1/2<J_1/J<1$, the ground state is four-fold degenerate. Thus we have a two-state orbital degree 
of freedom, for which operators can be represented by another set of Pauli spin matrices, which we denote $\tau$ matrices. In first order degenerate perturbation theory, we obtain the effective Hamiltonian:
\begin{equation}\label{eq6}
H_{eff}^1 = A \sum_{ij}(1+\tau_i^z\tau_j^z)\vec{S}_i\cdot\vec{S}_j,
\end{equation}
\begin{figure}[h]
\includegraphics[width=1.0\columnwidth]{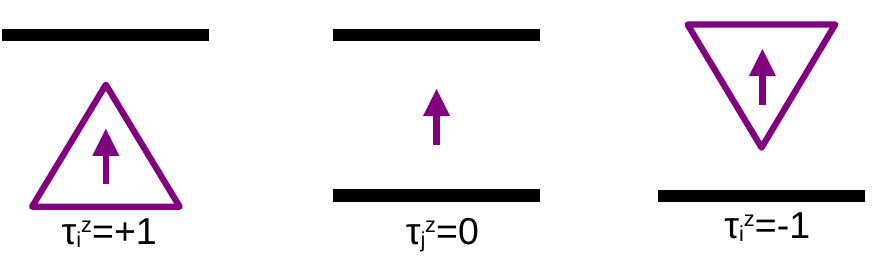}
\caption{Three orbitals states. ${\tau^z=+1,0,-1}$ are shown from left to right. Phase-III only has two orbital states $\tau^z_i=\pm 1$ at each site, whereas the effective Kugel-Khomskii model for phase-IV requires all 3 orbital states.}  
\label{fig:definition_of_orbitals}
\end{figure}
with $A=\frac{2 J_2}{9}$ (See supplementary materials for more details).
Here $\tau$ and $S$ are orbital and spin operators. $\tau_i^z$ is normalized to have eigenvalues $\pm 1$ (See also Fig.\ref{fig:definition_of_orbitals}) and $S_i^z$ has eigenvalues $\pm 1/2$. 

The expression $(1+\tau_i^z\tau_j^z)$ can take the value $0$ or $2$ for antiferromagnetic or ferromagnetic alignment of orbital degree. Therefore for positive $J_2$ ground state favours the ferromagnetic ordering of the orbitals and antiferromagnetic coupling of spins. The spin system behaves as spin-$1/2$ AF chain in presence of ferromagnetic orbital ordering. The ground state remains two-fold degenerate due to ferromagnetic orbital order while the spin excitations remains gapless with power-law spin correlations. 

In first order perturbation theory, there are no quantum fluctuations for the orbital degrees of freedom. However, this is no longer true
in second order degenerate perturbation theory. Second order perturbation theory generates new terms that allows quantum fluctuations of both orbitals and spins (See Supplementary Materials for more details). However, within phase-III, the broken symmetry associated with ferromagnetic orbital order or reflection around the horizontal axis remains. We will not discuss this case further in more detail as this perturbation theory does not provide insights outside phase-III.


We now focus on the special case of $J_1=1$.
From the phase diagram in Fig.~\ref{fig:phase_diagram}, it is clear that a degenerate perturbation theory only around this point can provide insights into phase-IV. For $J_2=0$ and $J_1=J$ the unit cell has six fold degeneracy with effective spin-$1/2$ and the unit cell configurations are shown in Fig.~\ref{fig:definition_of_orbitals}.  There are three orbital states in each unit cell. We continue to call the orbital operators as $\tau$. Note that they are no longer Pauli spin matrices but rather $3\times 3$ matrices. We denote the lower triangular state with $\tau^z=+1$, upper triangular state with $\tau^z=-1$ and central spin state with $\tau^z=0$. Degenerate perturbation theory in $J_2$ leads to the effective Hamiltonian,
\begin{equation}\label{eq7}
H_{eff}^1 = J_2 \sum_{ij} H_{ij}^\tau\  S_i\cdot S_j,
\end{equation}
where the orbital part of the operator is:
\begin{multline} \label{eq7}
    H_{ij}^\tau = \frac{4P_2}{9} - \frac{P_1}{3} (\tau_i^+\tau_j^- + \tau_i^-\tau_j^+) - \frac{1}{3} \Bar{P_1}(\tau_i^+\tau_j^+ + \tau_i^-\tau_j^-)\Bar{P_1}\\ - \frac{2}{3\sqrt{3}} (\ P_2( \tau_i^+ - \tau_j^+)P_1
    +P_1(\tau_i^- - \tau_j^-)P_2\ ).
    \end{multline}
Here, $\tau_i^\pm $ are orbital raising and lowering operators defined such that they raise/lower the $\tau^z$ value at site i by one unit and their non-zero matrix elements are unity. $P_2$ is a projection operator on to states with $\tau_i^z+\tau_j^z=\pm 2$, $P_1$ is a projection operator on to states with $\tau_i^z+\tau_j^z=\pm 1$ and $\overline{P_1}=1-P_1$. Thus under degenerate perturbation theory, the model maps on to a 1D Kugel-Khomskii model with a spin-one orbital degree of freedom with a large anisotropy in the orbital space. The $SU(2)$ spin symmetry remains.

The first term in Eq.~\ref{eq7} is the only term in common with the previous effective Hamiltonian in Eq.~\ref{eq6}. By itself it would lead to ferromagnetic orbital order as discussed before in phase-III. However, in phase-IV the other terms help to restore the Ising symmetry of the orbitals by strongly mixing the upper triangular states with the lower ones. In particular the third term mixes both upper and lower triangular states ($\tau^z=\pm1$) with the central spin states ($\tau^z=0$). 

The last terms are particularly interesting as they contain the seeds of dimerization and breaking of left-right symmetry within a five-site unit cell. They allow admixing of different orbital states on a single unit-cell without changing the basis states outside the unit cell and the term changes sign between the left and right unit-cells of the interaction. The net result is that interaction between effective sites $i$ and $i+1$ causes the central spin to become strongly correlated with the leg spins on the left for unit-cell $i$ and on the right for unit-cell $i+1$. This is one of the key features of the dimerization pattern in the phase-IV.

\begin{figure}[h]
\includegraphics[width=1.0\columnwidth]{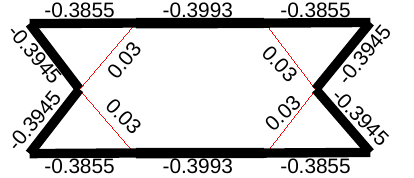}
\caption{Correlation functions, $C(r) = <S_i \cdot S_j>$ for 2 unit-cell cluster for the effective Hamiltonian in Eq.~\ref{eq7}.} 
\label{fig:correlation_effective_hamiltonian}
\end{figure}

We have checked that the effective Hamiltonian accurately reproduces the ground state energies and correlation functions of the full Hamiltonian for small values of $J_2$. The correlation functions for a two-site cluster with open boundary conditions with $J_2=0.01$ are shown in Fig.\ref{fig:correlation_effective_hamiltonian}. The key feature to notice is the lifting of left-right symmetry within a single unperturbed unit cell. The correlations are strengthened on the outside rungs and weakened on the inside rungs of a unit cell. 

\begin{figure*}[t]
\includegraphics[width=1.0\linewidth]{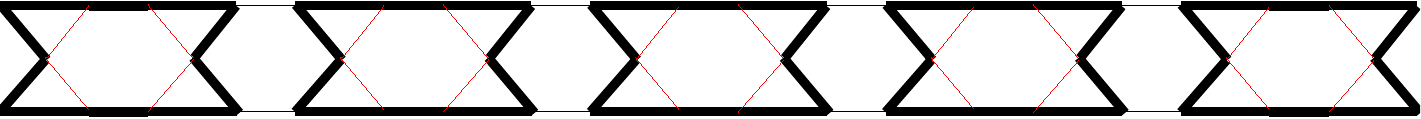}
\caption{Pattern of correlation functions, $C(r) = <S_i \cdot S_j>$ for the 10 unit-cell cluster with open boundary conditions from the effective Hamiltonian in Eq.~\ref{eq7} showing the doubled unit cell. The pattern of rung-bonds is the same as for $J_2=J_1$ but the correlations on the leg bonds varies widely within the phase.}
\label{fig:correlation_effective_hamiltonian_10}
\end{figure*}

As one goes to larger system sizes, we find that the open boundary condition suffices to break translational symmetry of the system and dimerize it. The entire system forms a repeating array of strongly correlated decagons each extending over two unit cells. 
Rung correlations by themselves form parts of alternating strong and weak hexagons as seen in Fig.~\ref{fig:correlation_effective_hamiltonian_10}. Note that translating the correlation pattern by one unit cell does not change the strength of intra-dimer leg bonds as it merely exchanges the left and right parts of the symmetric decagon. This explains why dimer order on leg bonds inside the unit cell $B_{14}^{01}$ and $B_{25}^{01}$ vanish exactly in this phase. The intra-unit cell broken symmetry on the rung-bonds and vanishing of $B_{14}^{01}$ and $B_{25}^{01}$ are key features of dimerization that persists throughout phase-IV. While the exact pattern of dimerization, especially on the legs of the KSC, evolves strongly within the phase with changes in exchange parameters, the character and symmetry of the phase remains unchanged.


\section{Summary}\label{conclusion}
In this paper, we have studied the quantum phase diagram of an isotropic spin-1/2 Heisenberg model with variation in exchange parameters on the Kagome strip chain system. 
We first  analysed the ground states of a unit cell by varying  $J_1/J$.  For $0 < J_1/J < 1/2$, the ground state has spin-$3/2$. For $1/2 < J_1/J < 1$, the ground state has two sets of $S=1/2$ states, which we denoted as spin-half state with an orbital degree of freedom (SHSODF). For $J_1/J=1$,  the ground state has $S=1/2$ with three orbital states.   For $J_1/J>1 $, the ground state is again a $S=1/2 $ doublet but with no additional degeneracy. 

With $J_2/J\neq 0$, we found  five distinct ground-state quantum phases.  Phases I and II have singlet ground states but have gapless spin-spectrum. These phases are related by a crossover of local properties. In both phases, the spin correlations are quasi-long-range ordered as in the Heisenberg chain. The difference in the two phases lies in the short-distance properties, which gradually changes with $J_2$ from a spin-$3/2$ in each unit cell for small $J_2$ which are weakly coupled from cell to cell, to one where the all the four leg spins in a unit cell are tied into singlets with spins in neighboring unit cells leaving only the central spin-$1/2$ in each unit cell to form quasi-long range order. The nearest-neighbor spin correlations evolve from being stronger within a unit cell to becoming stronger between unit cells. The bipartite entanglement entropy for a single unit cell evolves from $2\ln{2}$ when the spin in the unit cell is $3/2$ to an entropy of $5\ln{2}$ when all five of the spins in a unit cell entangle independently with spins outside the unit cell.

In phase-III the ground state is  doubly  degenerate. The degeneracy corresponds to ferromagnetic orbital ordering. The spin system remains power-law correlated with gapless spin excitations.  In  phase-IV, the ground state is a doubly degenerate singlet state which is dimerized with a resonating structure of dimer and hexamer pairs. The ground state has short range magnetic order  and a finite spin gap. This ground state was obtained earlier by  S. White et al\cite{RRP_Singh} using the DMRG calculations for the special case of $J_1/J=J_2/J=1$.  

In phase-V the ground state is a non-degenerate singlet and has quasi-long-range and short range   magnetic order for the central spins and  the leg spins  respectively.   The leg spin pairs  form strong intra-unit cell dimers in the ground state and  the  spin-1/2 at central sites forms an effective Heisenberg chain structure. The leg spin-pair singlets are only weakly perturbed for finite  $J_2$.  

To understand the phase-III and especially IV we carried out a degenerate  perturbation theory calculation in $J_2$  and constructed an effective Hamiltonian which resembles  a Kugel-Khomskii model. In phase-III, this model has two degenerate ground states each with ferromagnetically ordered orbitals together with an isotropic antiferromagnetic  spin-1/2 Heisenberg chain.  Phase-IV is the only phase with a spin-gap. It is a dimerized phase with a doubled unit cell. This is the phase that includes the uniformly coupled Kagome Strip chain $J_2=J_1=J$ studied previously. Perturbatively, this phase is related to a Kugel-Khomskii model with an anisotropic 3-state or spin-one orbital degree of freedom. In the small $J_2$ limit, the primary dimerization happens inside a unit cell, where the left and right rung correlations between leg and central spins become unequal and this pattern alternates from unit cell to unit cell. The detailed pattern of dimerization changes within the phase-IV but the essential character of the broken symmetry remains unchanged. In some parameter regimes, in this phase we find several singlet states below the lowest triplet state, a characteristic feature of the Kagome Lattice Heisenberg model. 

\section{Acknowledgement}
MK acknowledges support from SERB through Grant Sanction No. CRG/2020/000754. SG expresses appreciation for financial support from DST-INSPIRE. SG thanks Sudip Kumar Saha and Santanu Pal for the fruitful discussions. MK thanks Matthias Vojta for his suggestions.

\bibliography{ref}
\end{document}


\title{Frustrated spin-1/2 Heisenberg model on a Kagome-strip chain: Dimerization and mapping to a spin-orbital Kugel-Khomskii model}

\begin{abstract}
We here provide supplemental explanations and data on the following topics in relation to the main text: (I) Energetics of AFM kagome strip chain unit cell, Details of calculations of (II) degenerate perturbation analysis for $1/2<J_1 < 1$ and (III) $J_1 = 1$, (IV) Entanglement Entropy and (V) Spin-Spin correlation functions in different phases.
\end{abstract}

\author{Sayan Ghosh}
\affiliation{S.N. Bose National Centre for Basic Sciences, Kolkata 700098, India.}
\author{Rajiv R. P. Singh}
\email{rrpsing@ucdavis.edu}
\affiliation{Department of Physics, University of California Davis, Davis, California 95616, USA}
\author{Manoranjan Kumar}
\email{manoranjan.kumar@bose.res.in}
\affiliation{S.N. Bose National Centre for Basic Sciences, Kolkata 700098, India.}
\maketitle

\section{\label{SM1}Hamiltonian and its solution of kagome strip chain's unit cell}
The Hamiltonian for a unit cell can be written as,
\begin{equation} \label{eq1}
H_{uc}=J\Vec{S}_3.(\Vec{S}_1+\Vec{S}_2+\Vec{S}_4+\Vec{S}_5)+J_1(\Vec{S}_1.\Vec{S}_4+\Vec{S}_2.\Vec{S}_5). 
\end{equation}
$S_{14}$, $S_{25}$, $S_{1245}$ and $S_{12345}$ are all good quantum numbers. So, we can write the energy eigenvalue in the following form,
\begin{multline}\label{eq2}
E=\frac{J}{2}(\vec{S}_{t}^2-\vec{S}_3^2-\vec{S}_{1245}^2) + \frac{J_1}{2}(\vec{S}_{14}^2-\vec{S}_1^2-\vec{S}_4^2+\vec{S}_{25}^2-\vec{S}_2^2-\vec{S}_5^2)\\
 =\frac{J}{2}[s_t(s_{t}+1) - s_3(s_3+1) - s_{1245}(s_{1245}+1)] + \frac{J_1}{2}[s_{14}(s_{14}+1) + s_{25}(s_{25}+1) - s_1(s_1+1) - \\ s_2(s_2+1) - s_4(s_4+1) - s_5(s_5+1)]
\end{multline}
For spin- $1/2$ system, we can replace $s_1=s_2=s_3=s_4=s_5=1/2$ in Eq. \ref{eq2} and get the final form as following, 
\begin{equation}\label{eq3}
E= \frac{J}{2}[s_{t}(s_{t}+1) - \frac{3}{4} - s_{1245}(s_{1245}+1)] + \frac{J_1}{2}[s_{14}(s_{14}+1) + s_{25}(s_{25}+1) - 3]
\end{equation}
All the spin quantum number $s_{14}$, $s_{25}$, $s_{1245}$, $s_{t}$ and their corresponding thirty-two eigenvalues are listed in the following table.   
\begin{table}
    \centering
    \begin{tabular}{|c|c|c|c|c|c|} \hline 
         $s_{14}$&  $s_{25}$&  $s_{1245}$&  $s_{t}$&  $E$&  $s_t(s_t+1)$\\ \hline
 \rowcolor{green}1& 1& 2& 5/2& $J+J_1/2$& 6\\\hline \hline
         \rowcolor{lightgray}
         1&  1&  2&  3/2&  $-3J/2 + J_1/2$ &  4\\ \hline
         \rowcolor{lightgray}
         0&  1&  1&  3/2&  $J/2 - J_1/2$&  4\\ \hline
         \rowcolor{lightgray}
         1&  0&  1&  3/2&  $J/2 - J_1/2$&  4\\ \hline
         \rowcolor{lightgray}
         1&  1&  1&  3/2&  $J/2 + J_1/2$&  4\\ \hline 
         \rowcolor{yellow}
         0&  1&  1&  1/2&  $-J - J_1/2$&  2\\ \hline
         \rowcolor{yellow}
         1&  0&  1&  1/2&  $-J - J_1/2$&  2\\ \hline
         \rowcolor{yellow}
         1&  1&  1&  1/2&  $-J + J_1/2$&  2\\ \hline
         \rowcolor{yellow}
         0&  0&  0&  1/2&  $-3J_1/2$&  2\\ \hline
         \rowcolor{yellow}
         1&  1&  0&  1/2&  $J_1/2$&  2\\\hline
    \end{tabular}
    \caption{All possible combinations of $s_{14}$, $s_{25}$, $s_{1245}$ and $s_{t}$ are listed above. By putting those numbers in Eq. \ref{eq3}, energy eigenvalues have been calculated. All 32 states are shown in the table. $s_{t}=5/2$ arises only once and is never the ground state, $s_t=3/2$ arises $4$ times but only one can be a ground state, while $s_{t}=1/2$ arises $5$ times and multiple of them can form simultaneous ground states.}
    \label{energy_table}
\end{table}

\section{\label{SM2} Perturbation Analysis for Phase-III $1/2<J_1 < 1$}

In this regime ground state is four fold degenerate due to orbital and spin degrees of freedom (refer to the section III.B in main text). Four ground state wavefunctions are given by, $\ket{\tau^z=+1, S^z=+1/2}$, $\ket{\tau^z=+1, S^z=-1/2}$, $\ket{\tau^z=-1, S^z=+1/2}$, $\ket{\tau^z=-1, S^z=-1/2}$. Their expressions are provided below,
\begin{equation}\label{eq4}
\begin{split}
\ket{\tau^z=+1 \hspace{.3cm} S^z=+1/2} = \frac{1}{\sqrt{3}} (\ket{\uparrow\uparrow\downarrow\downarrow\uparrow} - \ket{\downarrow\uparrow\downarrow\uparrow\uparrow}) - \frac{1}{\sqrt{12}} (\ket{\uparrow\uparrow\uparrow\downarrow\downarrow} - \ket{\downarrow\uparrow\uparrow\uparrow\downarrow}) - \frac{1}{\sqrt{12}} (\ket{\uparrow\downarrow\uparrow\downarrow\uparrow} - \ket{\downarrow\downarrow\uparrow\uparrow\uparrow})\\
\ket{\tau^z=+1 \hspace{.3cm} S^z=-1/2} = \frac{1}{\sqrt{3}} (\ket{\uparrow\downarrow\uparrow\downarrow\downarrow} - \ket{\downarrow\downarrow\uparrow\uparrow\downarrow}) - \frac{1}{\sqrt{12}} (\ket{\uparrow\downarrow\downarrow\downarrow\uparrow} - \ket{\downarrow\downarrow\downarrow\uparrow\uparrow}) - \frac{1}{\sqrt{12}} (\ket{\uparrow\uparrow\downarrow\downarrow\downarrow} - \ket{\downarrow\uparrow\downarrow\uparrow\downarrow})
\end{split}    
\end{equation}
Other two ground state for $\tau^z = -1$ can be written in similar fashion. From these two we can evaluate $1^{st}$ order effective Hamiltonian. \\

\textit{\textbf{$1^{st}$ order perturbation theory:}}
Due to spin-rotational symmetry the Effective Hamiltonian in first order perturbation theory is:
\begin{equation}\label{eq5}
\mathcal{H}_{eff}^1 = J_2\sum_{ij}(1+\tau_i^z\tau_j^z)(A\vec{S}_i\cdot\vec{S}_j+B)
\end{equation}

In order to evaluate $A$ and $B$, we have to calculate matrix elements from two different basis, original spin basis of five site KSC unit cell $S$-basis and effective spin-orbital basis $\tau$-$S$ basis. In spin basis, original Hamiltonian contains $J_2(\Vec{S}_4^{(a)}.\Vec{S}_1^{(b)} + \Vec{S}_5^{(a)}.\Vec{S}_2^{(b)})$ for two unit cells $a$ and $b$. 

Now, in $S$-basis two matrix elements are calculated,

\begin{equation}\label{eq6}
\begin{split}
\bra{+1\hspace{.1cm}-1/2; \hspace{.3cm} +1\hspace{.1cm}-1/2} \Vec{S}_4^{(a)}\cdot\Vec{S}_1^{(b)} + \Vec{S}_5^{(a)}\cdot\Vec{S}_2^{(b)} \ket{+1\hspace{.1cm}-1/2; \hspace{.3cm} +1\hspace{.1cm}-1/2} = \frac{1}{9} \\
\bra{+1\hspace{.1cm}+1/2; \hspace{.3cm} +1\hspace{.1cm}-1/2} \Vec{S}_4^{(a)}\cdot\Vec{S}_1^{(b)} + \Vec{S}_5^{(a)}\cdot\Vec{S}_2^{(b)} \ket{+1\hspace{.1cm}+1/2; \hspace{.3cm} +1\hspace{.1cm}-1/2} = -\frac{1}{9}
\end{split}    
\end{equation}
Now above matrix elements of Eq.\ref{eq6} can be calculated in spin-orbital basis also.
\begin{equation}\label{eq7}
\begin{split}
\bra{+1\hspace{.1cm}-1/2; \hspace{.3cm} +1\hspace{.1cm}-1/2} \mathcal{H}_{eff}^1 \ket{+1\hspace{.1cm}-1/2; \hspace{.3cm} +1\hspace{.1cm}-1/2} = \frac{A}{2} + 2B \\
\bra{+1\hspace{.1cm}+1/2; \hspace{.3cm} +1\hspace{.1cm}-1/2} \mathcal{H}_{eff}^1 \ket{+1\hspace{.1cm}+1/2; \hspace{.3cm} +1\hspace{.1cm}-1/2} = -\frac{A}{2} + 2B
\end{split}    
\end{equation}
By comparing Eq. \ref{eq6} and \ref{eq7} we obtain, $A = \frac{2}{9}$ and $B = 0$. So, final effective Hamiltonian for $1^{st}$ order correction is reduced to,
\begin{equation}\label{eq8}
\mathcal{H}_{eff}^1 = \frac{2 J_2}{9}\sum_{ij}(1+\tau_i^z\tau_j^z)\vec{S}_i\cdot\vec{S}_j.
\end{equation}

All matrix elements involving the operators $\tau^+$ and $ \tau^-$ are absent in first order perturbation theory. This is because a transition between $\tau^z=+1$ and $\tau^z=-1$ in any unit cell requires changing the state of the spins on both upper and lower legs of the unit cell. However, since the perturbing Hamiltonian either has spin operators in the upper or lower leg such an operator cannot arise. This first order effective Hamiltonian has Ising orbital degrees of freedom with no orbital fluctuations. Thus the ground states have ferromagnetic orbital order in the ground state and the spin system reduces to a Heisenberg spin-half chain. 
\begin{figure}
    \centering
    \includegraphics[width=1.0\columnwidth]{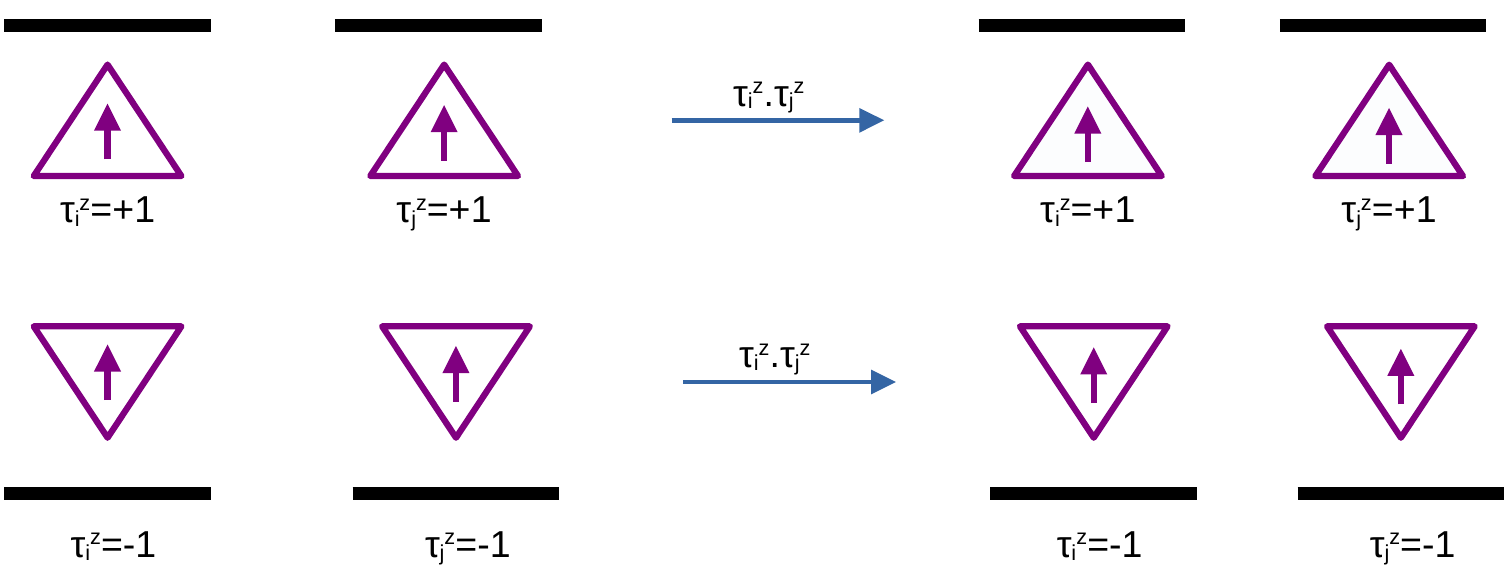}
    \caption{Operation of $\tau_i^z\tau_j^z$ on $\tau_i^z=+1,\tau_j^z=+1$ or $\tau_i^z=-1,\tau_j^z=-1$ for spin-$+1/2$. In similar way operation can be performed on spin-$-1/2$ also. The state will be unaltered. It is giving ferromagnetic ordering in orbital space.}
    \label{supfig:zz-orbital_space}
\end{figure}\\

\textit{\textbf{$2^{nd}$ order perturbation theory:}} In second-order perturbation theory the effective Hamiltonian is given by
$$
\mathcal{H}_{eff}^2 =\sum_{n,n \ne gs} \frac{\bra{\psi_{gs}} H' \ket{\psi_{n}}  \bra{\psi_{n}} H' \ket{\psi_{gs}} }{E_{gs}-E_{n}}.
$$
where $H'$ is the perturbing Hamiltonian and $\ket{\psi_{n}}$ and $\ket{\psi_{gs}}$ are states in the excited state and ground state manifolds respectively. Here, one can generate various additional terms in the effective Hamiltonian involving $\tau^+$ and $\tau^-$ operators by going through intermediate excited states corresponding to the $\tau^z=0$ (See Fig.~1 and Fig.~2 below) and \textbf{$s^z = 3/2$ states}. New operators in the effective Hamiltonian include $(\tau^+\tau^- + \tau^-\tau^+)$, $(\tau^+\tau^+ + \tau^-\tau^-)$ as well as $\tau^+$ and $\tau^-$ terms. We will discuss these processes in more detail in the next section where perturbation theory for $J_1/J=1$ is discussed, in which case $\tau^z=0$ states form part of the unperturbed ground state manifold and hence such processes arise already in first order perturbation theory. The spin part of the Hamiltonian must remain of the Heisenberg type in second order perturbation theory.

Second order perturbation theory gives rise to quantum fluctuations for the orbital degrees of freedom, which are no longer purely
Ising type. Thus we do not have two separate orbital ordered ground states but rather in a finite system the ground state becomes a superposition of various orbital states. However, as long as $J_2/J$ is small, second order terms are much smaller than first order and we stay in the same phase of ferromagnetic orbital order. 

Given that a large number of terms arise in second order, calculating the full second order perturbation theory requires a computational approach and we defer such studies of higher order perturbation theory to future work. 

\section{\label{SM3} Perturbation Analysis at $J_1=1$}
At $J_1=1$, the spin of the ground state is $1/2$ and it is three fold degenerate. Three fold degeneracy is equivalent to three fold orbital degree of freedom. From \ref{energy_table} we can find three possible combinations, $s_{14}=s_{25}=s_{1245}=0$, $s_{t}=1/2$; $s_{14}=1$, $s_{25}=0$, $s_{1245}=1$, $s_{t}=1/2$ and $s_{14}=0$, $s_{25}=1$, $s_{1245}=1$, $s_{t}=1/2$. They are pictorially depicted in the main text. Three orbitals are denoted as, $\ket{\tau^z}=\ket{-1}, \ket{0}, \ket{+1}$ and two effective spins are possible as total spin, $S=1/2$, $\ket{S^z}=\ket{+1/2}, \ket{-1/2}$. So, six degenerate ground states are possible- $\ket{-1 \hspace{.1cm} +1/2}$, $\ket{-1 \hspace{.1cm} -1/2}$, $\ket{0 \hspace{.1cm} +1/2}$, $\ket{0 \hspace{.1cm} -1/2}$, $\ket{+1 \hspace{.1cm} +1/2}$ and $\ket{+1 \hspace{.1cm} -1/2}$. Due to spin rotational symmetry, only $A\Vec{S}_i.\Vec{S}_j$ term is allowed in spin-orbital Hamiltonian.

\begin{figure}[h!]
    \centering
    \includegraphics[width=1.0\columnwidth]{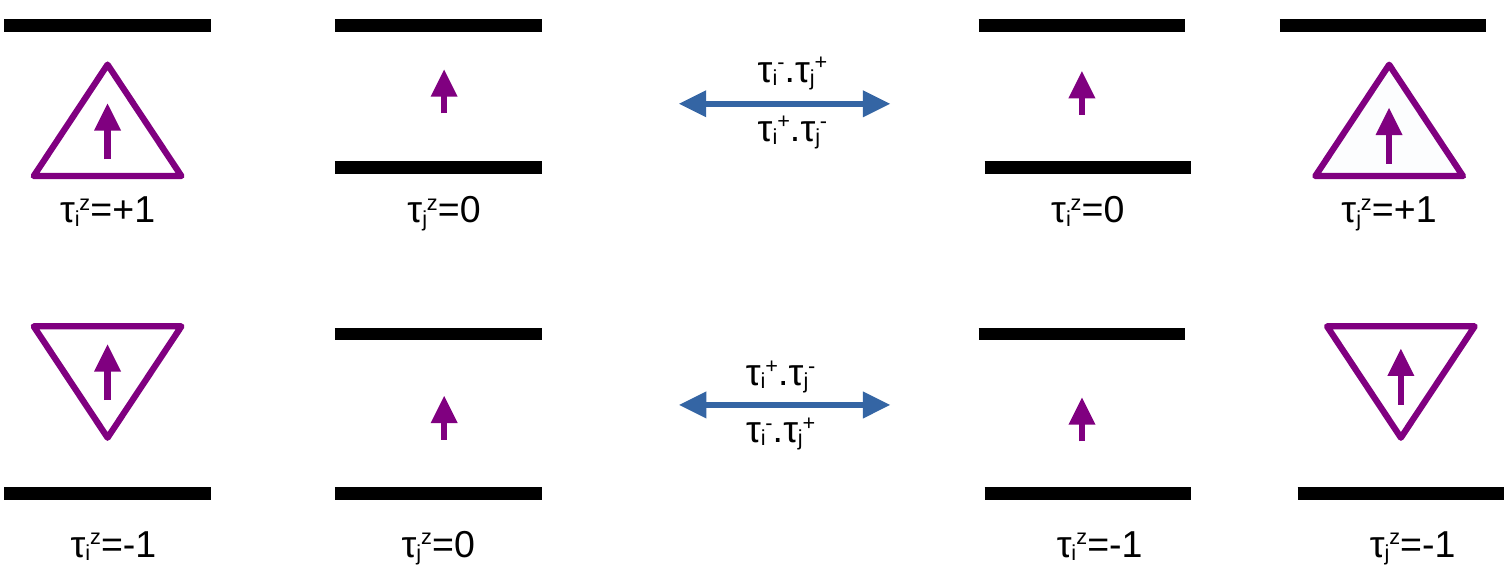}
    \caption{Operation of $\tau_i^+\tau_j^-$ or $\tau_i^-\tau_j^+$ on $\tau_i^z=+1,\tau_j^z=0$ or $\tau_i^z=-1,\tau_j^z=0$ shifts them to $\tau_i^z=0,\tau_j^z=+1$ or $\tau_i^z=0,\tau_j^z=+1$, respectively for spin-$+1/2$. In similar way operation can be performed on spin-$-1/2$ also.}
    \label{supfig:pm-mp_orbital}
\end{figure}

We define three different kind of Projection operators, $P_2$, $P_1$ and $\bar{P_1}$. Their operations are defined in the following for two sites:
\begin{equation}\label{eq9}
\begin{split}
P_2 \ket{+1 \hspace{.1cm} +1} = \ket{+1 \hspace{.1cm} +1}\\
P_2 \ket{-1 \hspace{.1cm} -1} = \ket{-1 \hspace{.1cm} -1}\\
P_2 \ket{all \hspace{.1cm} other \hspace{.1cm} states} = 0
\end{split}
\end{equation}
$P_2$ projector confined ourselves into $\tau_1^z + \tau_2^z = \pm 2$.
\begin{equation}\label{eq10}
\begin{split}
    P_1 \ket{+1 \hspace{.1cm} 0} = \ket{+1 \hspace{.1cm} 0}\\
    P_1 \ket{-1 \hspace{.1cm} 0} = \ket{-1 \hspace{.1cm} 0}\\
    P_1 \ket{0 \hspace{.1cm} +1} = \ket{0 \hspace{.1cm} +1}\\
    P_1 \ket{0 \hspace{.1cm} -1} = \ket{0 \hspace{.1cm} -1}\\
    P_1 \ket{all \hspace{.1cm} other \hspace{.1cm} states} = 0
\end{split}
\end{equation}
$P_1$ projector confined ourselves into $\tau_1^z + \tau_2^z = \pm 1$ and $\bar{P_1} = 1 - P_1$.
We have four types of terms in our Hamiltonian.
(i) From previous section we get $1^{st}$ term which is purely Heisenberg type.
\begin{equation}\label{eq11}
H_{eff}^1 = \frac{2 J_2}{9}\sum_{ij}(1+\tau_i^z\tau_j^z)\vec{S}_i\cdot\vec{S}_j
\end{equation}
As our $\tau_i^z + \tau_j^z = \pm 2$, that's why only spin exchange survives. Our Hamiltonian reduces to, 
\begin{equation}\label{eq12}
H_{eff1}^1 = \frac{4 J_2}{9}\sum_{ij}\Vec{S}_i.\cdot\Vec{S}_jP_2
\end{equation}
(ii) The above process (shown in Fig.\ref{supfig:pm-mp_orbital}) leads us to the following Hamiltonian,
\begin{equation}\label{eq12}
H_{eff2}^1 = P_1(\tau_i^+\tau_j^- + \tau_i^-\tau_j^+)(A\Vec{S}_i\cdot\Vec{S}_j) 
\end{equation}
We will calculate the matrix element in the original spin- $S$-basis on different states which are allowed by defined projector and then equate with the matrix element of effective spin-orbital basis to determine the value of $A$ in the \ref{eq12} Hamiltonian. By taking two sites we calculate matrix elements in $S$-basis.

\begin{figure}
    \centering
    \includegraphics[width=1.0\columnwidth]{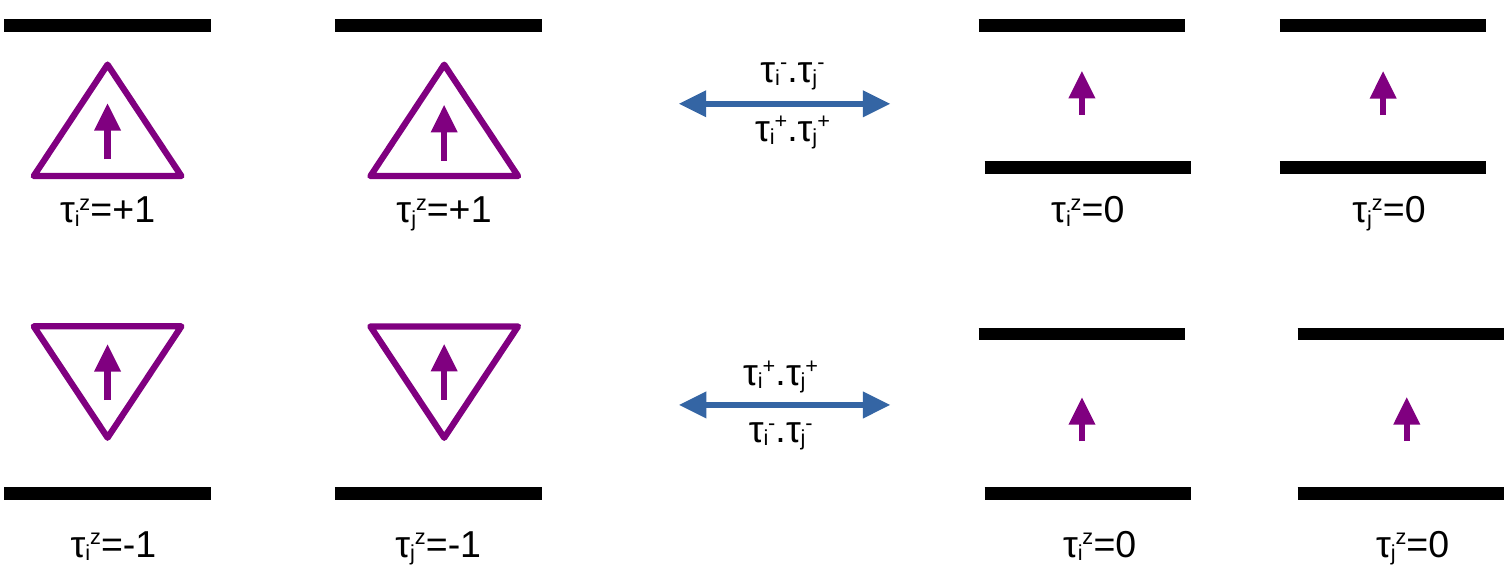}
    \caption{Operation of $\tau_i^+\tau_j^+$ or $\tau_i^-\tau_j^-$ on $\tau_i^z=+1,\tau_j^z=+1$ or $\tau_i^z=-1,\tau_j^z=-1$ shifts them to $\tau_i^z=0,\tau_j^z=0$ or $\tau_i^z=0,\tau_j^z=0$, respectively for spin-$+1/2$. In similar way operation can be performed on spin-$-1/2$ also.}
    \label{supfig:pp-mm_orbital}
\end{figure}

\begin{equation}\label{eq13}
\begin{split}
    \bra{+1 \hspace{.1cm} +1/2; \hspace{.2cm} 0 \hspace{.1cm} +1/2}\Vec{S}_4^{(a)}\cdot\Vec{S}_1^{(b)} + \Vec{S}_5^{(a)}\cdot\Vec{S}_2^{(b)} \ket{0 \hspace{.1cm} +1/2; \hspace{.2cm} +1 \hspace{.1cm} +1/2} = -\frac{1}{12}\\
    \bra{+1 \hspace{.1cm} +1/2; \hspace{.2cm} 0 \hspace{.1cm} -1/2} \Vec{S}_4^{(a)}\cdot\Vec{S}_1^{(b)} + \Vec{S}_5^{(a)}\cdot\Vec{S}_2^{(b)} \ket{0 \hspace{.1cm} +1/2; \hspace{.2cm} +1 \hspace{.1cm} -1/2} = -\frac{1}{12}
\end{split}    
\end{equation}
Now in spin-orbital basis corresponding matrix element is following,
\begin{equation}\label{eq14}
\begin{split}
    \bra{+1 \hspace{.1cm} +1/2; \hspace{.2cm} 0 \hspace{.1cm} +1/2}H_{eff2}^1\ket{0 \hspace{.1cm} +1/2; \hspace{.2cm} +1 \hspace{.1cm} +1/2} = \frac{A}{4}\\
    \bra{+1 \hspace{.1cm} +1/2; \hspace{.2cm} 0 \hspace{.1cm} -1/2}H_{eff2}^1\ket{0 \hspace{.1cm} +1/2; \hspace{.2cm} +1 \hspace{.1cm} -1/2} = \frac{A}{4}
\end{split}    
\end{equation}
From Eq.\ref{eq13} and \ref{eq14} we can get $A = -\frac{1}{3}$. So, final form of $H_{eff2}^1$ is reduced to,
\begin{equation}\label{eq15}
H_{eff2}^1 = -\frac{J_2}{3}\sum_{ij} P_1 (\tau_i^+\tau_j^- + \tau_i^-\tau_j^+) \vec{S}_i \cdot \vec{S}_j
\end{equation}
(iii) The final process (shown in Fig.\ref{supfig:pp-mm_orbital}) leads us to the following Hamiltonian,
\begin{equation}\label{eq16}
H_{eff3}^1 = \Bar{P_1}(\tau_i^+\tau_j^+ + \tau_i^-\tau_j^-)\Bar{P_1}(B\Vec{S}_i \cdot \Vec{S}_j) 
\end{equation}
Similarly we calculate matrix elements in $S$-basis and effective spin orbital basis
\begin{equation}\label{eq17}
\begin{split}
    \bra{+1 \hspace{.1cm} +1/2; \hspace{.2cm} +1 \hspace{.1cm} +1/2}\Vec{S}_4^{(a)}\cdot\Vec{S}_1^{(b)} + \Vec{S}_5^{(a)}\cdot\Vec{S}_2^{(b)} \ket{0 \hspace{.1cm} +1/2; \hspace{.2cm} 0 \hspace{.1cm} +1/2} = -\frac{1}{12}\\
    \bra{+1 \hspace{.1cm} +1/2; \hspace{.2cm} +1 \hspace{.1cm} -1/2}\Vec{S}_4^{(a)}\cdot\Vec{S}_1^{(b)} + \Vec{S}_5^{(a)}\cdot\Vec{S}_2^{(b)}\ket{0 \hspace{.1cm} +1/2; \hspace{.2cm} 0 \hspace{.1cm} -1/2} = -\frac{1}{12}
\end{split}    
\end{equation}
Now in spin-orbital basis corresponding matrix element is following,
\begin{equation}\label{eq18}
\begin{split}
    \bra{+1 \hspace{.1cm} +1/2; \hspace{.2cm} +1 \hspace{.1cm} +1/2}H_{eff3}^1\ket{0 \hspace{.1cm} +1/2; \hspace{.2cm} 0 \hspace{.1cm} +1/2} = \frac{B}{4}\\
    \bra{+1 \hspace{.1cm} +1/2; \hspace{.2cm} +1 \hspace{.1cm} -1/2}H_{eff3}^1\ket{0 \hspace{.1cm} +1/2; \hspace{.2cm} 0 \hspace{.1cm} -1/2} = \frac{B}{4}
\end{split}    
\end{equation}
By comparing Eq.\ref{eq17} and \ref{eq18} we can get $B = -\frac{1}{3}$. So, final form of $H_{eff3}^1$ is reduced to,
\begin{equation}\label{eq19}
H_{eff3}^1 = -\frac{J_2}{3}\sum_{ij}\Bar{P}_1(\tau_i^+\tau_j^+ + \tau_i^-\tau_j^-)\Bar{P}_1 \Vec{S}_i \cdot \Vec{S}_j 
\end{equation}
(iv) There is one more matrix element. From $S$-basis we can evaluate the matrix element as follows,
\begin{equation}\label{eq21}
\begin{split}
\bra{+1\hspace{.1cm}+1/2; \hspace{.3cm} 0\hspace{.1cm}+1/2} \Vec{S}_4^{(a)}\cdot\Vec{S}_1^{(b)} + \Vec{S}_5^{(a)}\cdot\Vec{S}_2^{(b)} \ket{+1\hspace{.1cm}+1/2; \hspace{.3cm} +1\hspace{.1cm}+1/2} = -\frac{1}{6\sqrt{3}} \\
\bra{0\hspace{.1cm}+1/2; \hspace{.3cm} +1\hspace{.1cm}+1/2} \Vec{S}_4^{(a)}\cdot\Vec{S}_1^{(b)} + \Vec{S}_5^{(a)}\cdot\Vec{S}_2^{(b)} \ket{+1\hspace{.1cm}+1/2; \hspace{.3cm} +1\hspace{.1cm}+1/2} = \frac{1}{6\sqrt{3}}
\end{split}    
\end{equation}
The same matrix element can be retrieved from $\tau-S$ basis. The corresponding Hamiltonian can be expressed as,
\begin{equation}\label{eq22}
H_{eff4}^1 = C [P_2 (\tau_i^+ - \tau_j^+) P_1  + P_1 (\tau_i^- - \tau_j^-) P_2] \Vec{S}_i \cdot \Vec{S}_j 
\end{equation}
So, corresponding matrix element in $\tau-S$ basis are as following,
\begin{equation}\label{eq23}
\begin{split}
    \bra{+1\hspace{.1cm}+1/2; \hspace{.3cm} 0\hspace{.1cm}+1/2}H_{eff4}^1\ket{+1\hspace{.1cm}+1/2; \hspace{.3cm} +1\hspace{.1cm}+1/2} = \frac{C}{4}\\
    \bra{0\hspace{.1cm}+1/2; \hspace{.3cm} +1\hspace{.1cm}+1/2}H_{eff4}^1\ket{+1\hspace{.1cm}+1/2; \hspace{.3cm} +1\hspace{.1cm}+1/2} = -\frac{C}{4}
\end{split}    
\end{equation}
By comparing Eq.\ref{eq21} and \ref{eq23} we get,
\begin{equation}\label{eq22}
H_{eff4}^1 = -\frac{2}{3\sqrt{3}} [P_2 (\tau_i^+ - \tau_j^+) P_1  + P_1 (\tau_i^- - \tau_j^-) P_2] \Vec{S}_i \cdot \Vec{S}_j 
\end{equation}
Now the final form of effective Hamiltonian (upto $1^{st}$ order correction) is given by,
\begin{equation}\label{eq25}
\begin{split}
\mathcal{H}_{eff}^1 = \frac{4 J_2}{9}\sum_{ij}P_2\Vec{S}_i\cdot\Vec{S}_j -\frac{J_2}{3}\sum_{ij}P_1(\tau_i^+\tau_j^- + \tau_i^-\tau_j^+)P_1 \Vec{S}_i.\Vec{S}_j  - \frac{J_2}{3}\sum_{ij}\Bar{P}_1(\tau_i^+\tau_j^+ + \tau_i^-\tau_j^-)\Bar{P}_1 \Vec{S}_i.\Vec{S}_j \\
- \frac{2}{3\sqrt{3}} \sum_{ij} [P_2 (\tau_i^+ - \tau_j^+) P_1  + P_1 (\tau_i^- - \tau_j^-) P_2] \Vec{S}_i \cdot \Vec{S}_j
\end{split}
\end{equation}

\section{\label{SM4}Entanglement Entropy}
We study the von-Neumann entropy to understand the entangled states in the unit cell with rest of the system. Assuming the unit cell as system and rest of the system as environment, the total Hilbert space can be given as, $H_{AB} = H_A \otimes H_B$, where $H_A$ and $H_B$ are the Hilbert spaces associated with $A$ and $B$. If $\rho_{AB}$ is the density matrix of the composite system, then the reduced density matrices of A and B are defined as follows : 
\begin{equation} \label{eq4}
\begin{split}
\rho_A = \text{tr}_B(\rho_{AB}), \\
\rho_B = \text{tr}_A(\rho_{AB}).
\end{split}
\end{equation}
Here, $\text{tr}_A$ and $\text{tr}_B$ represent the partial traces over the degrees of freedom of A and B, respectively. The entropy of the entanglement of formation can be expressed as:

\begin{equation} \label{eq4}
\begin{split}
S(\rho_A) = -\text{tr}(\rho_A \log \rho_A) = S(\rho_B) = -\text{tr}(\rho_B \log \rho_B).
\end{split}
\end{equation}

In our specific case, we construct the reduced density matrix for a unit cell (where 20 sites are divided into two parts - unit cell A of five sites and environment B of fifteen sites) and measure the entanglement entropy. The density matrix is given by $\rho_{AB} = \ket{\psi_{gs}}\bra{\psi_{gs}}$ if $\psi_{gs}$ is non-degenerate.

\begin{figure}[h]
\includegraphics[width=1.0\columnwidth]{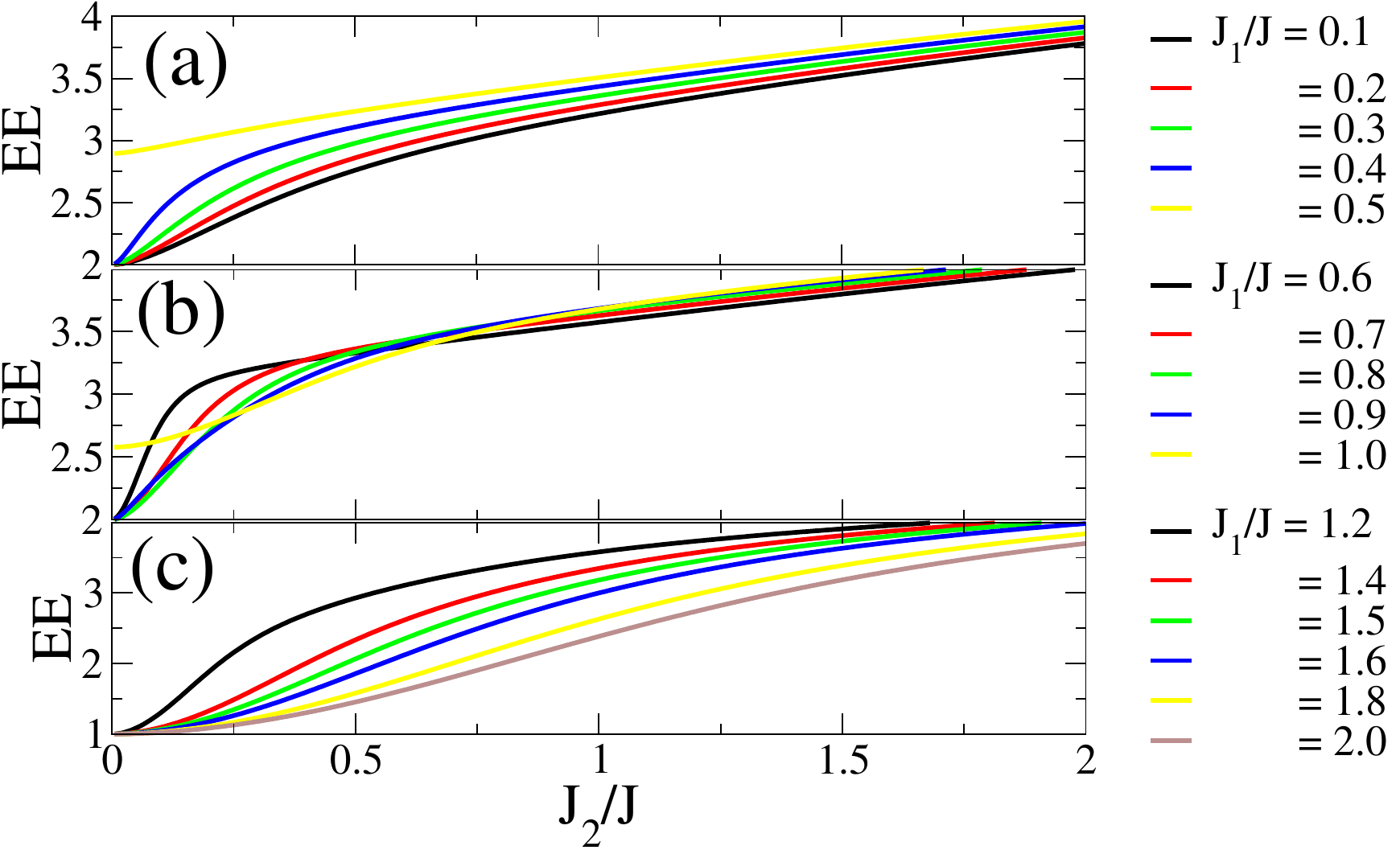}
\caption{Entanglement entropy (EE) as a function of $J_2/J$ for a fixed $J_1/J$, measured in units of $\ln 2$. (a) In phase I, the EE is approximately $2$ at $J_2 \approx 0$, indicating the presence of four degrees of freedom. As $J_2/J$ increases, the degrees of freedom within the unit cell also increase. (b) In phase III, the EE remains $2$ at $J_2 \approx 0$, suggesting four degrees of freedom. (c) In phase V, the EE is $1$ at $J_2 \approx 0$, corresponding to two degrees of freedom.}
\label{fig:entanglement entropy}
\end{figure}

\section{\label{SM5}Spin-Spin Correlation}
\begin{figure}[h!]
\includegraphics[width=1.0\columnwidth]{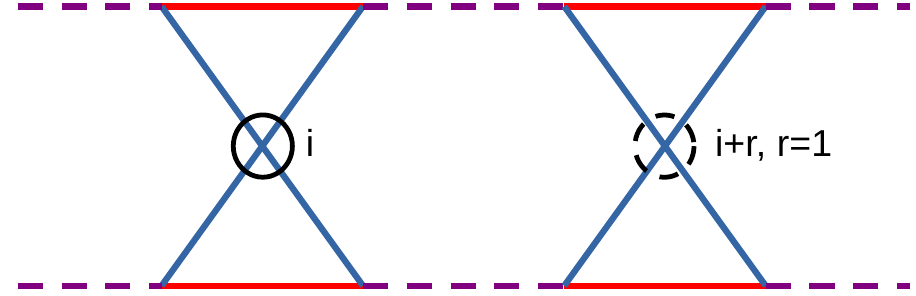}
\caption{Structure of kagome strip chain for showing reference site and distance from reference. Black solid circle shows reference site $i$ and broken black circle shows the unit distance $i+r$, $r=1$.}
\label{fig:ref_distance}
\end{figure}

We have measured spin-spin correlation, $C(r) = \bra{\psi_0} S_i \cdot S_{i+r} \ket{\psi_0}$, where reference of the central site $i$ and the distance $r$ between the two central sites are shown in Fig.\ref{fig:ref_distance}. We calculate the correlation function in the system with $32$ unit cells with open boundary condition. Reference site $i$ is always chosen at the middle of the system. 

\begin{figure}[h!]
\includegraphics[width=1.0\columnwidth]{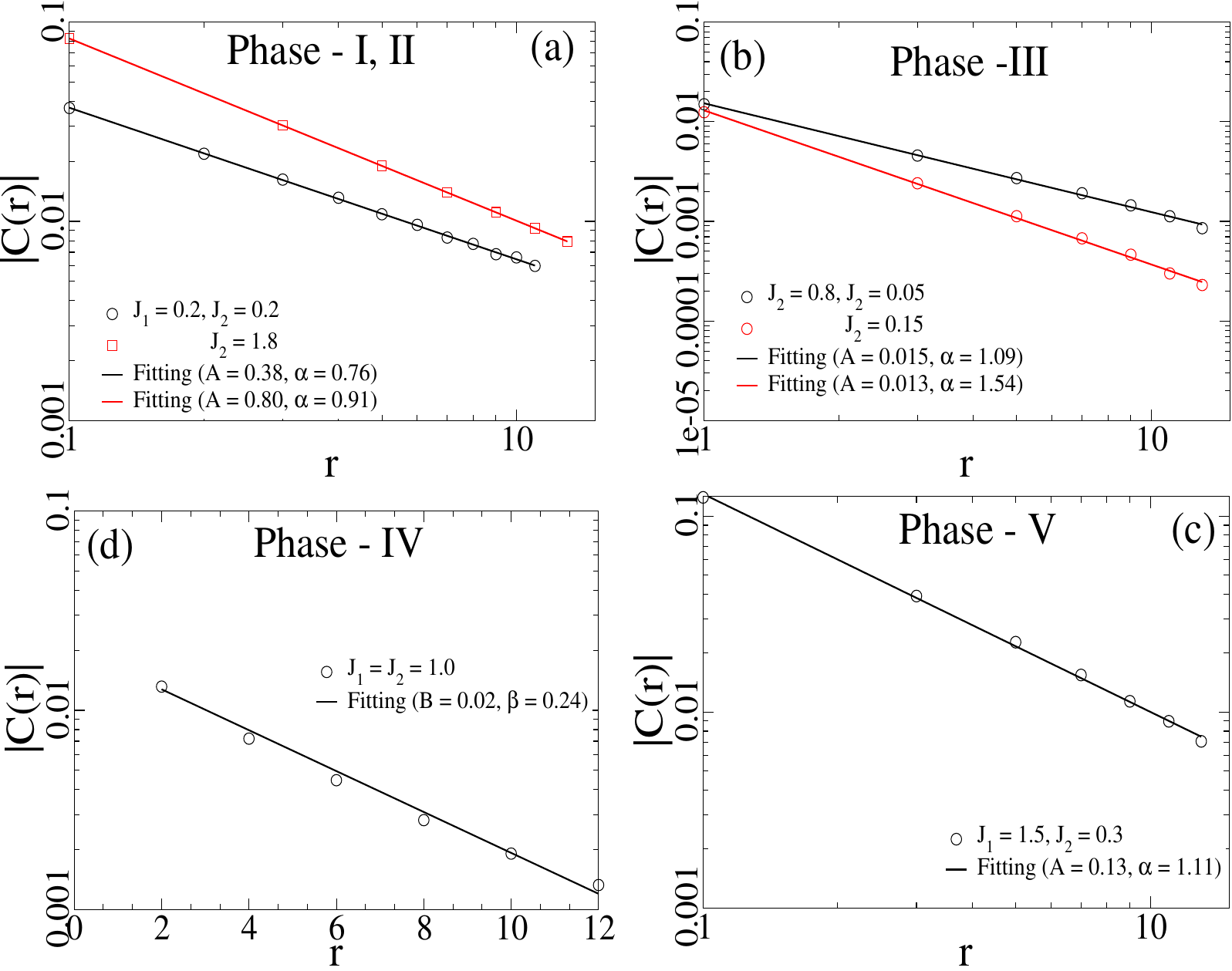}
\caption{Correlation function $C(r)$ is plotted in (a) phase-I,II, (b) phase-III, (c) phase-V where it shows power law behaviour ($Ar^{-\alpha}$) and (d) phase-IV where it decays exponentially ($Bexp(-\beta r)$). $\alpha$ and $\beta$ provides the information about how fast correlation decays. Fittings are shown by solid lines and symbols represent the calculated correlation functions.}
\label{fig:correlation_decay}
\end{figure}

In Fig.\ref{fig:correlation_decay}(a,b,c), we notice that the absolute value of correlation functions between the central spin follow power law ($Ar^{-\alpha}$) in phase-I, II, III and V, whereas it decays exponentially ($Bexp(-\beta r)$) in phase-IV as shown in Fig.\ref{fig:correlation_decay}(d).

In Table \ref{table:correlation_decay} we summarize the nature of correlation function between central spins in all the five phases.

\begin{table}[H]
    \caption{Nature of correlation decay in different phases}
\label{table:correlation_decay}
\centering
\setlength{\tabcolsep}{9.5pt}
    \begin{tabular}{c   c   c}    
        Phase&  Nature of correlation decay& Ordering based upon correlation decay\\
        \hline
         I&  Power& Quasi Long Range\\ 
         II&  Power& Quasi Long Range\\ 
         III&  Power& Quasi Long Range\\ 
         IV&  Exponential& Short Range\\ 
         V&  Power& Quasi Long Range\\ 
        \hline
        \hline
    \end{tabular}
\end{table}